\documentclass{article}
\usepackage{amsfonts}
\usepackage{amsmath}

\input{tcilatex}

\begin{document}

\title{Contribution from stochastic electrodynamics to the understanding of
quantum mechanics}
\author{L. de la Pe\~{n}a\thanks{%
Corresponding author. Email: luis@fisica.unam.mx} \ and A. M. Cetto$\thanks{%
Present address: Department of Technical Cooperation, IAEA. Wagramer Strasse
5, A-1400 Vienna, Austria.}$ \\
Instituto de F\'{\i}sica, Universidad Nacional Aut\'{o}noma de M\'{e}xico\\
Apartado Postal 20-364, M\'{e}xico 01000, M\'{e}xico}
\date{}
\maketitle

\begin{abstract}
During the last decades there has been a relatively extensive attempt to
develop the theory of stochastic electrodynamics (\textsc{sed}) with a view
to establishing it as the foundation for quantum mechanics. The theory had
several important successes, but failed when applied to the study of
particles subject to nonlinear forces. An analysis of the failure showed
that its reasons are not to be ascribed to the principles of \textsc{sed},
but to the methods used to construct the theory, particularly the use of a
Fokker-Planck approximation and perturbation theory. A new, non perturbative
approach has been developed, called linear stochastic electrodynamics (%
\textsc{lsed}), of which a clean form is presented here.

After introducing the fundamentals of \textsc{sed}, we discuss in detail the
principles on which \textsc{lsed} is constructed. We pay attention to the
fundamental issue of the mechanism that leads to the quantum behaviour of
field and matter, and demonstrate that indeed \textsc{lsed} is a natural way
to the quantum formalism by demanding its solutions to comply with a limited
number of principles, each one with a clear physical meaning. As a further
application of the principles of \textsc{lsed} we derive also the Planck
distribution. In a final section we revisit some of the most tantalizing
quandaries of quantum mechanics from the point of view offered by the
present theory, and show that it offers a clear physical answer to them.

\end{abstract}

\section{Introduction}

Despite the extraordinary power of quantum mechanics, it is difficult to
find in the history of physics another example of a theory that has raised
and nourished so many debates and controversies about its meaning. The
myriads of papers, books and meetings devoted to the scrutiny of its
interpretation testify to the meager progress reached in such disputes since
the early stages of quantum mechanics. Of course for the physicist who uses
the theory in her/his daily undertakings as a tool, there is usually nothing
to bother about, so she/he will easily overlook such questions. But there
exist also some physicists (not so few as to be negligible, as evidenced by
the number of papers) who are deeply interested in the study and solution of
these matters and would relish a clear-cut answer to them. Being
simultaneously a fundamental theory and an active field of research, quantum
theory cannot fully flourish while indefinitely leaving aside the basic
conceptual issues that are known to affect it. Thus no wonder that the
papers dealing with the fundamentals of quantum theory continue to
accumulate, as is easily confirmed by a glance, for example, at the recent
monograph by Auletta \cite{Aul}. Most frequently the problem is tackled from
within the quantum theory itself, as is well illustrated with the works of
Bohm \cite{Bohm} or Omn\`{e}s \cite{Omnes}. In the course of time, however,
several attempts to find a solution to those problems from a broader
framework have been developed, as testified in the old book by Jammer \cite%
{Jam} or, of course, the more recent one by Auletta.

Among the varied efforts to construct a theory aimed at contributing to the
understanding of quantum mechanics, we shall refer here specifically to 
\emph{stochastic electrodynamics}, \textsc{sed} for short. (For its origins
see the pioneering works of T. W. Marshall \cite{Mar63}, who has contributed
also to its optical branch, \emph{stochastic optics }\cite{MarSanVid94}, 
\cite{MarSan02}). We recall that the central premise of \textsc{sed} is that
the quantum behavior of the particle is a result of its interaction with the
vacuum radiation field, or zero-point field. This field is assumed to
pervade the space and, for the purpose of studying atomic or molecular
systems, is considered to be in a stationary state with well defined
stochastic properties. Its action on the particle impresses upon it in every
point of space a stochastic motion, with an intensity characterized by
Planck's constant, which is a measure of the magnitude of the fluctuations
of the vacuum field.

Initially put forward by Nernst as a conjecture, the crucial role of
vacuum-matter interaction for the stability of the atom and other quantum
properties is a fundamental result in \textsc{sed} (see \cite{Dice} for an
almost exhaustive list of references to the end of 1995). Phenomena as
diverse and as characteristic of the quantum world as van der Waals and
Casimir forces \cite{Boy70c}, \cite{Boy75a}, diamagnetism \cite{Mar63}, \cite%
{Boy80a}, \cite{dlPJau83}, cavity effects on atomic systems \cite{cavity},
thermal effects of acceleration \cite{Boy84}-\cite{Rue90}, the quantum
harmonic oscillator including its radiative corrections \cite{San74}-\cite%
{FraMar88}, and several others, have found a consistent, even if in some
cases still incomplete physical explanation within the framework of \textsc{%
sed}, as can be seen in the detailed account given in \cite{Dice}. This
collection of fitting results suggested that the core of the theory is a
sound one. However, the difficulties encountered when applying it to systems
subject to nonlinear forces \cite{Boy76}-\cite{San85} brought the theory
almost to a standstill, except for the renewed efforts by Cole and coworkers
to advance in the understanding of the hydrogen atom \cite{Col90}, \cite%
{ColZou03}, the development of stochastic optics by Marshall, Santos and
coworkers \cite{MarSanVid94}, \cite{MarSan02}, and the proposal of an
alternative formulation of \textsc{sed} by Cetto and de la Pe\~{n}a \cite%
{Dice}, \cite{elaf}, \cite{dlPCet01}. This latter is the subject matter of
the present paper.

This alternative theory, termed \emph{linear stochastic electrodynamics} (%
\textsc{lsed}) for reasons that will become clear below, shares with \textsc{%
sed} its basic principle about the central role played by the zero-point
field, but makes a careful review of supplementary hypotheses used along the
derivation of the theory. A detailed analysis as the one given in \cite{Dice}
convinced us that the culprit was to be found in the use of a perturbative
method to deal with the effect of the field on the mechanical system, and
the corresponding use of a generalized Fokker-Planck equation \cite{note0}.
This has prompted us to propose a nonperturbative approach that makes no
recourse to methods associated with the Fokker-Planck treatment of
stochastic problems \cite{Dice}, \cite{elaf}. In the course of time we have
been able to refine the arguments that sustain this approach, and we believe
that the framework presented here constitutes a more accomplished form of 
\textsc{lsed,} suitable for those physicists who would like to see quantum
mechanics emerge as a physical theory devoid of strange elements or
assumptions. So here we show how the usual formalism of quantum mechanics
(and quantum electrodynamics, in its nonrelativistic version) ensues from a
more general underlying theory, and we use this demonstration to understand
the origin of some major quantum peculiarities.

It is not our intention to offer here a full derivation of quantum theory
but to provide the fundamental elements that explain how the usual quantum
formalism comes about. As will be apparent below, the nature of \textsc{lsed}
is such as to expect that it should lead us beyond the framework of present
day quantum theory. This certainly constitutes a most attractive feature of
the theory that should encourage its further development. For the time being
we take the limited step of applying the principles of the theory to
understand the physics of today; but already along this restricted way we
shall come across some novelties.

The paper is organized as follows. Using as starting point the
Abraham-Lorentz equation of motion for a particle subject to an arbitrary
external force and the zero-point field, in the first sections we set forth
three principles that limit to a considerable extent the class of allowed
solutions, by imposing to them clearly defined statistical demands. We
explicitly state the approximations introduced to satisfy each one of the
principles. We then approach the problem of finding these solutions and show
that they are naturally described by the formalism of (nonrelativistic)
quantum theory. Having reached that point we take a closer look at the
interaction of matter with the radiation field, to distinguish between this
equilibrium field and the vacuum field in the absence of matter, i.e., the
free vacuum field. As an application we study the equilibrium with a thermal
field (including the vacuum, of course) to derive Planck's distribution
along the lines of the old statistical method proposed by Einstein \cite%
{Ein16}. The final section differs somewhat in nature from the rest of the
paper in that we use it to address, in the light of the present theory, some
of the conceptual problems of quantum mechanics that have been under
discussion for decades.

\section{The principles of LSED}

We consider the problem of a bound particle, typically an electron, and
start from the Abraham-Lorentz equation of \textsc{sed} (usually called
Braffort-Marshall equation in this context) 
\begin{equation}
m\overset{\cdot \cdot }{\mathbf{x}}=\mathbf{f}(\mathbf{x})+m\tau \overset{%
\cdot \cdot \cdot }{\mathbf{x}}+e\mathbf{E}(t).  \label{11}
\end{equation}%
The quantity $\tau =2e^{2}/3mc^{3}$ is of the order of 10$^{-23}$ s for the
electron. The radiation reaction force $m\tau \overset{\cdot \cdot \cdot }{%
\mathbf{x}}$ has the well known causality problems associated with the
third-order time derivative, but these are of no special concern to us here;
in \cite{Dice} the interested reader may find a detailed discussion of this
point and an ample list of relevant references. The term $e\mathbf{E}(t)$
stands for the electric force exerted by the vacuum field on the particle;
the magnetic term is not included since the discussion is restricted to the
nonrelativistic case. Further, as will become evident later, the wavelength
of the relevant field modes is assumed to be much larger than the amplitude
of the particle's motion. This allows us to neglect the spacial dependence
of $\mathbf{E}$; in other words, we are using the electric dipole
approximation.

\subsection{The quantum regime}

Let us now discuss in detail the premises on which \textsc{lsed} is based.
The first assumption to be made is the following.

\textbf{Principle\ One}. \emph{The system under study reaches an equilibrium
state, at which the average rate of energy radiated by the particle equals
the average rate of energy absorbed by it from the field}.

To give a quantitative form to this demand, we multiply Eq.(\ref{11}) by $%
\overset{\cdot }{\mathbf{x}}$ and get after some minor transformations%
\begin{equation}
\left\langle \frac{dH}{dt}\right\rangle =-m\tau \left\langle \overset{\cdot
\cdot }{\mathbf{x}}^{2}\right\rangle +e\left\langle \overset{\cdot }{\mathbf{%
x}}\cdot \mathbf{E}\right\rangle ,  \label{12}
\end{equation}%
where $H$ stands for the particle Hamiltonian, including the Schott energy,%
\begin{equation}
H=\tfrac{1}{2}m\overset{\cdot }{\mathbf{x}}^{2}+V(\mathbf{x})-m\tau \overset{%
\cdot }{\mathbf{x}}\cdot \overset{\cdot \cdot }{\mathbf{x}},  \label{13}
\end{equation}%
and $V(\mathbf{x})$ is the potential associated to the external force $%
\mathbf{f}(\mathbf{x}).$ The average is being performed over the
realizations of the background (zero-point) field. When the system has
reached the state of energetic equilibrium, so that 
\begin{subequations}
\begin{equation}
\left\langle \frac{dH}{dt}\right\rangle =0,  \label{wa}
\end{equation}%
we have 
\begin{equation}
m\tau \left\langle \overset{\cdot \cdot }{\mathbf{x}}^{2}\right\rangle
=e\left\langle \overset{\cdot }{\mathbf{x}}\cdot \mathbf{E}\right\rangle .
\label{wb}
\end{equation}%
The two sides of this equation are of a very different nature: energy
radiation, its average rate being given by the Larmor term $m\tau
\left\langle \overset{\cdot \cdot }{\mathbf{x}}^{2}\right\rangle ,$ is due
basically to the orbital motions, whereas energy absorption, whose average
rate is $e\left\langle \overset{\cdot }{\mathbf{x}}\cdot \mathbf{E}%
\right\rangle , $ comes from the highly irregular motion impressed on the
particle by the vacuum field, and more specifically from the radiative
(stochastic) corrections to the primary motions, as will be argued below
(see Eq.(\ref{100})). When this equilibrium condition is satisfied (or
nearly satisfied) we say that the system has reached the \emph{quantum regime%
}. The theory to be developed assumes that this regime has been reached.
Below we will show that in \textsc{lsed} an even more stringent condition is
satisfied, namely that of detailed energy balance, i.e., balance for each
separate frequency. Of course this is to be expected under equilibrium,
since otherwise energy could be transferred by the mechanical part of the
system from some modes of the field to others, in clear violation of the
principles of thermodynamics.

\subsection{Central role of the vacuum field}

Despite the fact that Eq.(\ref{wb}) is still unfinished (as shown by Eq.(\ref%
{100})) we can draw some initial conclusions from it. One of primary
importance is that in equilibrium, $\left\langle \overset{\cdot \cdot }{%
\mathbf{x}}^{2}\right\rangle $ is determined by the vacuum field (more
specifically, by its energy spectrum, as we will see below). Thus, also the
acceleration itself should be determined by the field. The importance of
this observation can be recognized by considering a counterexample. Suppose
that we examine a state of motion determined \emph{perturbatively} from Eq.(%
\ref{11}), taking the field as the perturbation. Then the dominant part of $%
\mathbf{x}$ comes from the classical equation of motion $m\overset{\cdot
\cdot }{\mathbf{x}}=\mathbf{f}(\mathbf{x})$ (along with the field we are
neglecting the radiation damping). Under these conditions there is no
guarantee that the radiated power $m\tau \left\langle \overset{\cdot \cdot }{%
\mathbf{x}}^{2}\right\rangle $ equals $e\left\langle \overset{\cdot }{%
\mathbf{x}}\cdot \mathbf{E}\right\rangle $ for each possible motion, since $%
\mathbf{f}(\mathbf{x})$ and $\mathbf{E(t)}$ are entirely independent
functions. This was precisely the problem created by the original form of 
\textsc{sed}, as is discussed in detail in \cite{Dice} and \cite{elaf}. From
this it follows that it was not \textsc{sed} itself which failed with the
nonlinear forces, but the approach developed to study it. We conclude that
there is a need to look for a different kind of solutions, such that the
acceleration is determined by the vacuum field and Eq.(\ref{wb}) is
guaranteed to hold for all allowed states of motion. We embody this
observation in the form of Principle Two:

\textbf{Principle Two}. \emph{Once the quantum regime has been attained (and
Eq}.(\ref{wb})\emph{\ holds), the vacuum field has gained control over the
motion of the material part of the system}.

To apply this principle to the present problem, we consider the equation of
motion (\ref{11}) in the first place for the free particle, 
\end{subequations}
\begin{equation}
m\overset{\cdot \cdot }{x}=m\tau \overset{\cdot \cdot \cdot }{x}+eE(t),
\label{20}
\end{equation}%
and to simplify the discussion we consider the one-dimensional case, as
there seems to be no problem in generalizing to the multidimensional
instance. Now we express the field as a Fourier transform as follows,%
\begin{equation}
E(t)=\sum_{\beta }\widetilde{E}_{\beta }a_{\beta }e^{i\omega _{\beta
}t}=\sum_{\omega _{\beta }>0}\left( \widetilde{E}_{\beta }^{(+)}a_{\beta
}e^{i\omega _{\beta }t}+\widetilde{E}_{\beta }^{(-)}a_{\beta }^{\ast
}e^{-i\omega _{\beta }t}\right) .  \label{21}
\end{equation}%
The amplitudes $a_{\beta }=a(\omega _{\beta })$ are stochastic variables
with statistical properties that will be fixed from the requirements of the
theory itself. In the usual form of \textsc{sed} it was customary to fix a
priori these properties by writing them in the form%
\begin{equation}
a(\omega _{\beta })=r(\omega _{\beta })e^{i\varphi (\omega _{\beta })},
\label{22}
\end{equation}%
with both $r$ and $\varphi $ real functions, the amplitude $r$ following a
normal distribution and the phases $\varphi (\omega _{\beta })$ being
independent random numbers uniformly distributed in $\left[ 0,2\pi \right] ,$
as corresponds to a \emph{free} field. Here we follow a different path and
leave the $a(\omega _{\beta })$ largely unspecified for the time being,
since we will find below that the statistical properties of the (near) \emph{%
equilibrium} field cannot be freely fixed, but must follow from the
principles of the theory. The field amplitudes $\widetilde{E}_{\beta }$ will
be selected so as to assign to each mode of the field the mean energy $%
\mathcal{E}_{\beta }=\frac{1}{2}\hbar \omega _{\beta }.$ This is the unique
door through which Planck's constant enters into the theory, fixing the
scale of the spectral energy of the zero-point field \cite{rho}; from here
it spreads over the whole theory. Thus we write%
\begin{equation}
\mathcal{E}_{\beta }=\tfrac{1}{2}\left\langle p_{\beta }^{2}+\omega _{\beta
}^{2}q_{\omega }^{2}\right\rangle ,  \label{22e1}
\end{equation}%
with%
\begin{equation}
p_{\beta }=\sqrt{\frac{\mathcal{E}_{\beta }}{2}}\left( a_{\omega }+a_{\omega
}^{\ast }\right) ,\quad i\omega _{\beta }q_{\beta }=\sqrt{\frac{\mathcal{E}%
_{\beta }}{2}}\left( a_{\omega }-a_{\omega }^{\ast }\right) .  \label{22e2}
\end{equation}%
With (\ref{21}) the solution to Eq.(\ref{20}) becomes 
\begin{subequations}
\begin{equation}
x(t)=\sum \widetilde{x}_{\beta }a_{\beta }e^{i\omega _{\beta }t},
\label{23a}
\end{equation}%
where 
\begin{equation}
\widetilde{x}_{\beta }=-\frac{e\widetilde{E}_{\beta }}{m\omega _{\beta
}^{2}+im\tau \omega _{\beta }^{3}}.  \label{23b}
\end{equation}%
Hence, 
\begin{equation}
x(t)=-\sum_{\beta }\frac{e\widetilde{E}_{\beta }a_{\beta }}{m\omega _{\beta
}^{2}+im\tau \omega _{\beta }^{3}}e^{i\omega _{\beta }t}.  \label{23c}
\end{equation}%
In these expressions all quantities except the amplitudes $a_{\beta }$ are
sure numbers. It is important to note that this includes the amplitudes $%
\widetilde{x}_{\beta }$ and the frequencies $\omega _{\beta }$. Upon
introduction of an external force $f(x)$, however, these parameters become
in principle stochastic variables. Indeed, from the complete equation of
motion (\ref{11}) we get 
\end{subequations}
\begin{equation}
\sum \left( -m\omega _{\beta }^{2}\widetilde{x}_{\beta }-im\tau \omega
_{\beta }^{3}\widetilde{x}_{\beta }+\frac{\widetilde{f}_{\beta }}{a_{\beta }}%
\right) a_{\beta }e^{i\omega _{\beta }t}=e\sum \widetilde{E}_{\beta
}a_{\beta }e^{i\omega _{\beta }t}.  \label{24}
\end{equation}%
For a generic force, the Fourier coefficient $\widetilde{f}_{\beta }$ (of
the terms that oscillate with frequency $\omega _{\beta }$) will be a
complicated function of both sets, $\left\{ \widetilde{x}_{\beta }\right\} $
and $\left\{ a_{\beta }\right\} .$ By writing 
\begin{subequations}
\begin{equation}
\widetilde{x}_{\beta }=-\frac{e\widetilde{E}_{\beta }}{m\omega _{\beta
}^{2}+im\tau \omega _{\beta }^{3}+\frac{\widetilde{f}_{\beta }}{\widetilde{x}%
_{\beta }a_{\beta }}}  \label{25a}
\end{equation}%
and introducing this into Eq.(\ref{23a}), we get 
\begin{equation}
x(t)=-\sum \frac{e\widetilde{E}_{\beta }a_{\beta }}{m\omega _{\beta
}^{2}+im\tau \omega _{\beta }^{3}+\frac{\widetilde{f}_{\beta }}{\widetilde{x}%
_{\beta }a_{\beta }}}e^{i\omega _{\beta }t}.  \label{25b}
\end{equation}%
Comparing with (\ref{23c}) we see that introduction of an external force
modifies the response amplitudes $\widetilde{x}_{\beta }$ in a way that may
be very important. Firstly, the $\widetilde{x}_{\beta }$ referring to
different frequencies become entangled, so that the response to a given
frequency depends now on the response to other frequencies; secondly, as
already noted they become stochastic parameters, functions of the field
amplitudes $a_{\beta },$ as is seen from Eq.(\ref{25a}). Also, the $x(t)$
given by Eq.(\ref{25b}) is determined in an essential way by both the field
and the external force, so that it fulfils Principle Two. It should further
be noted that one gets a different solution for each realization of the
vacuum field, i.e., for each set $\left\{ a_{\beta }\right\} $, so that for
nonlinear forces, when $\widetilde{f}_{\beta }$ becomes a nonlinear function
of the sets $\left\{ \widetilde{x}_{\beta }\right\} $ and $\left\{ a_{\beta
}\right\} $, we have a continuous infinity of stochastic solutions.

\subsection{Looking for solutions independent of the realization of the field%
}

The problem of determining $x(t)$ in the general case appears impossible to
solve. However there is a way to considerably simplify matters, to the
extent of transforming the problem into a soluble one, under certain
restrictions. Let us consider an ensemble of similarly prepared systems.
Owing to differences in the initial conditions, specific realizations of the
background field and so on, there would be a whole collection of different
states of motion. We are however interested in those that are particularly
stable, and which thus become dominant as equilibrium is approached, at
least in the mean. One expects that such particularly stable orbits would be
those corresponding to a minimum average energy in some appropriate
(statistical) sense (to be detailed below, see subsection \ref{stab}). Owing
to their greater stability, such motions will result approximately the same
(in the statistical sense just mentioned) for a whole family of realizations
of the field. Thus, one can characterize them by being to a certain extent
independent of the details of the field realization. We propose to stretch
this consideration to its limits and consider those motions in the quantum
regime that become independent of the realization of the background field.

This request is clearly equivalent to demanding that the near equilibrium
vacuum field has adjusted itself to the presence of matter in the given
state of motion. A similar situation takes place, for instance, with the
equilibrium field at a temperature $T$, which is not merely the vacuum
field, but that corresponding to the Planck distribution at the given
temperature. This is the reason we have formerly stated that the field
amplitudes $a_{\beta }$ should be fixed by the demand of equilibrium. The
price to be paid for the present major simplification is that the theory
becomes unable to describe the detailed behavior of any particular member of
the ensemble (or subensemble) considered. It is in this sense that the
description has become statistical. This stripped-down description is simply
accomplished noticing that the set of solutions described by Eq.(\ref{25a})
satisfies the stated demand when the amplitudes $a_{\beta }$ are such that $%
\widetilde{f}_{\beta }/\left( \widetilde{x}_{\beta }a_{\beta }\right) $
becomes independent of the specific realization.

From the above discussion it follows that the amplitudes $a_{\beta }$ should
be selected so as to guarantee that the following Principle Three holds.

\textbf{Principle Three}. \emph{There exist states of matter (quantum
states) that are unspecific to (or basically independent of) the particular
realization of the vacuum field}.

The demand that the system of Eqs.(\ref{25a})-(\ref{25b}) possesses
solutions that satisfy Principle Three will be considered as the simplest
possible approximation to the solutions that satisfy the condition (\ref{wb}%
) of (detailed) energy balance, that is, once the quantum regime has been
attained. Prior to this, the field may be anyone, possibly closer to the
free field. To establish the consequences of Principle Three we expand the
Fourier amplitude $\widetilde{f}_{\beta }$ of the external force that
corresponds to the frequency $\omega _{\beta }$ as follows, noting that each
factor $\widetilde{x}_{\beta }$ should carry an associated $a_{\beta }$
factor, as follows from Eq.(\ref{23a}) (we leave aside the case of a
constant force), 
\end{subequations}
\begin{subequations}
\begin{equation}
\widetilde{f}_{\beta }=k_{\beta }\widetilde{x}_{\beta }a_{\beta }+k_{\beta
^{\prime }\beta ^{\prime \prime }}\widetilde{x}_{\beta ^{\prime }}\widetilde{%
x}_{\beta ^{\prime \prime }}a_{\beta ^{\prime }}a_{\beta ^{\prime \prime
}}+k_{\beta ^{\prime }\beta ^{\prime \prime }\beta ^{\prime \prime \prime }}%
\widetilde{x}_{\beta ^{\prime }}\widetilde{x}_{\beta ^{\prime \prime }}%
\widetilde{x}_{\beta ^{\prime \prime \prime }}a_{\beta ^{\prime }}a_{\beta
^{\prime \prime }}a_{\beta ^{\prime \prime \prime }}+...  \label{28a}
\end{equation}%
The point in this expansion is that the nonlinear terms entangle the
frequencies, so that there may appear an arbitrary number of terms
associated with the same frequency. Each one of the factors $\widetilde{x}%
_{\beta ^{\prime }}$ is accompanied by the factor $a_{\beta ^{\prime
}}e^{i\omega _{\beta ^{\prime }}t},$ so that the product of the time
functions gives the factor $e^{i\left( \omega _{\beta ^{\prime }}+\omega
_{\beta ^{\prime \prime }}+\omega _{\beta ^{\prime \prime \prime }}+\ldots
\right) t}=e^{i\omega _{\beta }t},$ a fact that has been already taken into
account in writing Eq.(\ref{25b}). Thus it follows that 
\begin{equation}
\frac{\widetilde{f}_{\beta }}{\widetilde{x}_{\beta }a_{\beta }}=k_{\beta
}+k_{\beta ^{\prime }\beta ^{\prime \prime }}\frac{\widetilde{x}_{\beta
^{\prime }}\widetilde{x}_{\beta ^{\prime \prime }}}{\widetilde{x}_{\beta }}%
\frac{a_{\beta ^{\prime }}a_{\beta ^{\prime \prime }}}{a_{\beta }}+k_{\beta
^{\prime }\beta ^{\prime \prime }\beta ^{\prime \prime \prime }}\frac{%
\widetilde{x}_{\beta ^{\prime }}\widetilde{x}_{\beta ^{\prime \prime }}%
\widetilde{x}_{\beta ^{\prime \prime \prime }}}{\widetilde{x}_{\beta }}\frac{%
a_{\beta ^{\prime }}a_{\beta ^{\prime \prime }}a_{\beta ^{\prime \prime
\prime }}}{a_{\beta }}+...  \label{28b}
\end{equation}%
Now it is clear that the response functions $\widetilde{x}_{\beta }$ will
become sure numbers, independent of the field realization, if the set of
conditions 
\end{subequations}
\begin{equation}
a_{\beta ^{\prime }}a_{\beta ^{\prime \prime }}a_{\beta ^{\prime \prime
\prime }}...a_{\beta ^{(n)}}=a_{\beta }  \label{30}
\end{equation}%
is satisfied for any number of factors, since $\widetilde{f}_{\beta }$
reduces then to 
\begin{equation}
\widetilde{f}_{\beta }=\left( k_{\beta }\widetilde{x}_{\beta
}+\sum\nolimits^{\beta }k_{\beta ^{\prime }\beta ^{\prime \prime }}%
\widetilde{x}_{\beta ^{\prime }}\widetilde{x}_{\beta ^{\prime \prime
}}+\sum\nolimits^{\beta }k_{\beta ^{\prime }\beta ^{\prime \prime }\beta
^{\prime \prime \prime }}\widetilde{x}_{\beta ^{\prime }}\widetilde{x}%
_{\beta ^{\prime \prime }}\widetilde{x}_{\beta ^{\prime \prime \prime
}}+...\right) a_{\beta }  \label{31}
\end{equation}%
so that Eq.(\ref{25a}) acquires sure values, 
\begin{equation}
\widetilde{x}_{\beta }=-\frac{e\widetilde{E}_{\beta }}{m\omega _{\beta
}^{2}+im\tau \omega _{\beta }^{3}+k_{\beta }+\sum^{\beta }k_{\beta ^{\prime
}\beta ^{\prime \prime }}\frac{\widetilde{x}_{\beta ^{\prime }}\widetilde{x}%
_{\beta ^{\prime \prime }}}{\widetilde{x}_{\beta }}+\sum^{\beta }k_{\beta
^{\prime }\beta ^{\prime \prime }\beta ^{\prime \prime \prime }}\frac{%
\widetilde{x}_{\beta ^{\prime }}\widetilde{x}_{\beta ^{\prime \prime }}%
\widetilde{x}_{\beta ^{\prime \prime \prime }}}{\widetilde{x}_{\beta }}+...}.
\label{32}
\end{equation}%
It is to be noted that this expression is exact (although implicit) whenever
conditions (\ref{30}) are satisfied. However, since the amplitudes $a_{\beta
}$ are stochastic quantities it would be naive to assume that the latter are
satisfied exactly. So with the present approximations we are also neglecting
the ``noise''\ associated with all such fluctuations. As remarked above,
together with requirement (\ref{30}) a condition on the combination
frequencies must be satisfied, namely, 
\begin{equation}
\omega _{\beta ^{\prime }}+\omega _{\beta ^{\prime \prime }}+\omega _{\beta
^{\prime \prime \prime }}+\cdots +\omega _{\beta ^{(n)}}=\omega _{\beta }.
\label{33}
\end{equation}%
so that each term in the denominator of Eq.(\ref{32}) corresponds to the
common frequency $\omega _{\beta }.$ We call \emph{relevant frequencies} all
those frequencies that solve equation (\ref{33}); a central problem of the
theory will be their determination. Note that equation (\ref{33}) is weaker
than Eq.(\ref{30}): if the latter is met, the former will be automatically
satisfied, but not in the opposite sense. That condition (\ref{30}) is
satisfied is the meaning of the superscript $\beta $ in the sums $%
\sum\nolimits^{\beta }$ in the above equations.

Of course Eq.(\ref{32}) correctly contains the particular case of the
harmonic oscillator, for which $k_{\beta }=-m\omega _{0}^{2},$ $k_{\beta
^{\prime }\beta ^{\prime \prime }}=k_{\beta ^{\prime }\beta ^{\prime \prime
}\beta ^{\prime \prime \prime }}=...=0,$ just as a particular instance of
the general description. This is to be remarked because in the original
(conventional) \textsc{sed} theory, the harmonic oscillator, being a linear
system, was dealt with directly with a Fourier development, just as is done
here, so that the answer is the same in both theories, except for those
features that depend on the different statistical properties of the $%
a_{\beta }.$ However, in conventional \textsc{sed,} nonlinear problems are
treated using perturbation theory around the corresponding classical motion,
so that nothing equivalent to Eq.(\ref{32}) for the general case has place
in that theory. In the present theory, we have extended the treatment of the
harmonic oscillator to the generic case. There remains however an important
difference between the linear oscillator and the more general problem; this
comes from the fact that for the oscillator we get 
\begin{equation}
\widetilde{x}_{\beta }=-\frac{e\widetilde{E}_{\beta }/m}{\omega _{\beta
}^{2}+i\tau \omega _{\beta }^{3}-\omega _{0}^{2}},  \label{40}
\end{equation}%
so there is no explicit need to impose the conditions (\ref{30}), which
means that for all stochastic fields (or all realizations of a given field)
one obtains the same set $\widetilde{x}_{\beta }$. This includes the free
vacuum field (the one assumed in conventional \textsc{sed}), as well as
several other representations of the quantized radiation field. This is but
a manifestation of a well known result, namely, that the harmonic oscillator
can reach an equilibrium state with any background field. We could say that
what we have achieved here is equivalent to extending the property of the
linear harmonic oscillator of being independent of the specific realization
of the field, to all dynamical systems in the quantum regime.

\section{Solutions in the quantum regime}

Let us now attempt to give a precise meaning to the above equations and find
their solution. We start by considering conditions (\ref{30}) $a_{\beta
^{\prime }}a_{\beta ^{\prime \prime }}a_{\beta ^{\prime \prime \prime
}}...a_{\beta ^{(n)}}=a_{\beta },$ which we rewrite using the polar
representation (\ref{22}), $a_{\beta }=r_{\beta }e^{i\varphi _{\beta }}$, to
obtain%
\begin{equation}
r_{\beta ^{\prime }}r_{\beta ^{\prime \prime }}r_{\beta ^{\prime \prime
\prime }}...r_{\beta ^{(n)}}e^{i\left( \varphi _{\beta ^{\prime }}+\varphi
_{\beta ^{\prime \prime }}+\varphi _{\beta ^{\prime \prime \prime
}}+...+\varphi _{\beta ^{(n)}}\right) }=r_{\beta }e^{i\varphi _{\beta }}.
\label{41}
\end{equation}%
Since the number of factors $r_{\beta }$ in the left hand side is arbitrary
and their product should equal $r_{\beta }$ in all cases, this equation
requires that we take $r_{\beta }=1,$ and so on$.$ Thus the stochastic
amplitudes simplify to%
\begin{equation}
a_{\beta }=e^{i\varphi _{\beta }},  \label{42}
\end{equation}%
with $\varphi _{\beta }$ a random phase uniformly distributed in $\left[
-\pi ,\pi \right] .$ With this, equation (\ref{41}) reduces to%
\begin{equation}
\varphi _{\beta ^{\prime }}+\varphi _{\beta ^{\prime \prime }}+\varphi
_{\beta ^{\prime \prime \prime }}+...+\varphi _{\beta ^{(n)}}=\varphi
_{\beta }.  \label{43}
\end{equation}%
Thus the conditions on the phases and on the (relevant) frequencies, Eq.(\ref%
{33}) become similar. It is clear that Eqs.(\ref{33}) and (\ref{43}) relate
only relevant frequencies or phases among themselves; in other words, not
any phase (or frequency) enters into the conditions, so that our problem is
just the specification of the relevant ones. Let us consider first the case
of only two phases, so that 
\begin{subequations}
\begin{equation}
\varphi _{\beta ^{\prime }}+\varphi _{\beta ^{\prime \prime }}=\varphi
_{\beta }.  \label{44a}
\end{equation}%
It is clear that even if the phases entering in this equation are random,
they have become correlated one with another. So we may write for instance%
\begin{equation}
\varphi _{\beta ^{\prime }}=\varphi _{\beta }+\phi _{\beta ^{\prime }\beta
},\quad \varphi _{\beta ^{\prime \prime }}=\varphi _{\beta }+\phi _{\beta
^{\prime \prime }\beta },\quad \varphi _{\beta ^{\prime }}=\varphi _{\beta
^{\prime \prime }}+\phi _{\beta ^{\prime }\beta ^{\prime \prime }},
\label{44c}
\end{equation}%
or 
\end{subequations}
\begin{equation}
\phi _{\beta ^{\prime }\beta }=\varphi _{\beta ^{\prime }}-\varphi _{\beta
},\quad \text{and so on,}  \label{phase}
\end{equation}%
with each $\phi _{\beta ^{\prime }\beta ^{\prime \prime }}$ a random phase.
Substituting in Eq.(\ref{44a}) one obtains 
\begin{equation*}
\varphi _{\beta }+\phi _{\beta ^{\prime }\beta }+\varphi _{\beta ^{\prime
}}-\phi _{\beta ^{\prime }\beta ^{\prime \prime }}=\varphi _{\beta ^{\prime
\prime }}-\phi _{\beta ^{\prime \prime }\beta }+\phi _{\beta ^{\prime }\beta
}+\varphi _{\beta ^{\prime }}-\phi _{\beta ^{\prime }\beta ^{\prime \prime
}}=\varphi _{\beta },
\end{equation*}%
or, simplifying with the help of Eq.(\ref{44a}),%
\begin{equation}
\phi _{\beta ^{\prime }\beta }=\phi _{\beta ^{\prime }\beta ^{\prime \prime
}}+\phi _{\beta ^{\prime \prime }\beta }.  \label{45}
\end{equation}%
Since, according to Eq.(\ref{44c}), each phase $\phi _{\beta ^{\prime }\beta
}$ can be written as the difference of two random phases, and the latter are
uniformly distributed in $\left[ -\pi ,\pi \right] ,$ also the $\phi _{\beta
^{\prime }\beta }$ are uniformly distributed in the same interval, modulo $%
\pi $ \cite{Papo}. This result establishes the condition that the $\phi
_{\beta ^{\prime }\beta }$ should obey to guarantee that Eq.(\ref{44a}) is
satisfied. We thus find that the indices of the new phases must follow a
chain rule as shown in Eq.(\ref{45}), which is easily generalized to any
number of terms, so for the general case we have%
\begin{equation}
\phi _{\beta ^{\prime }\beta }=\phi _{\beta ^{\prime }\beta ^{\prime \prime
}}+\phi _{\beta ^{\prime \prime }\beta ^{\prime \prime \prime }}+\phi
_{\beta ^{\prime \prime \prime }\beta ^{\prime \prime \prime \prime
}}+\ldots +\phi _{\beta ^{(n-1)}\beta ^{(n)}}+\phi _{\beta ^{(n)}\beta }.
\label{46}
\end{equation}%
In terms of the original phases $\varphi _{\beta }$ the mechanism that leads
to the fulfilment of Eq.(\ref{43}) is the successive cancellation of pairs
of phases. Thus, for instance Eq.(\ref{46}) is equivalent to%
\begin{equation*}
\varphi _{\beta ^{\prime }}-\varphi _{\beta }\hspace{-0.06cm}=\hspace{-0.06cm%
}\varphi _{\beta ^{\prime }}\overset{0}{\overbrace{-\varphi _{\beta ^{\prime
\prime }}+\varphi _{\beta ^{\prime \prime }}}\,}\overset{0}{\overbrace{%
-\varphi _{\beta ^{\prime \prime \prime }}+\varphi _{\beta ^{\prime \prime
\prime }}}}\,\overset{0}{\overbrace{-\varphi _{\beta ^{\prime \prime \prime
\prime }}+\varphi _{\beta ^{\prime \prime \prime \prime }}}\,}-\ldots 
\overset{0}{\overbrace{-\varphi _{\beta ^{(n)}}+\varphi _{\beta ^{(n)}}}}%
-\varphi _{\beta },
\end{equation*}%
which is automatically satisfied. In summary, this means that Eq.(\ref{43})
should be written in terms of the phases $\phi _{\beta ^{\prime }\beta }$
instead of the original phases $\varphi _{\beta },$ and that in that doing
the original single index should be replaced by a pair of indices that
fulfil the chain rule. This also means that Eq.(\ref{42}) should be
rewritten in the form%
\begin{equation}
a_{\alpha \beta }=e^{i\phi _{\alpha \beta }}=e^{i\left( \varphi _{\alpha
}-\varphi _{\beta }\right) },  \label{50}
\end{equation}%
and, more generally, that the original single index should be replaced
throughout by a pair of indices that combine themselves according to the
chain rule made explicit in Eq.(\ref{46}). In particular, we must apply this
rule to Eq.(\ref{33}), $\omega _{\beta }=\omega _{\beta ^{\prime }}+\omega
_{\beta ^{\prime \prime }}+\omega _{\beta ^{\prime \prime \prime }}+\cdots
+\omega _{\beta ^{(n)}},$ which now reads 
\begin{equation}
\omega _{\beta \beta ^{(n)}}=\omega _{\beta \beta ^{\prime }}+\omega _{\beta
^{\prime }\beta ^{\prime \prime }}+\omega _{\beta ^{\prime \prime }\beta
^{\prime \prime \prime }}+\cdots +\omega _{\beta ^{(n-1)}\beta ^{(n)}}.
\label{51}
\end{equation}%
This is the precise meaning that one should ascribe to the symbol $%
\sum^{\beta }$ used in previous equations, as (\ref{31}) and (\ref{32}). The
frequencies that enter into all these relations are just the relevant
frequencies defined above, and the $a_{\alpha \beta }$ are the relevant
stochastic amplitudes.

It is clear that the demand (\ref{33}) has implied a drastic reduction of
``useful''\ frequencies and stochastic amplitudes to those that qualify as
relevant, leaving the rest aside from the present consideration. This is a
direct and most important consequence of Principle Three. Although the
``non-relevant''\ frequencies still exist and operate, their combined action
reduces merely to a noise that we have been systematically neglecting and
that adds to the motions described by the present approximation. Indeed they
are part of the source of the (nonrelativistic) radiative corrections, as
will become clear below. This explains also a most tantalizing feature of
the present theory, namely that the behavior of a mechanical (atomic) system
controlled by the random vacuum field may be described in terms of response
functions and characteristic (relevant) frequencies that are sure numbers.
As it is clear from the present discussion, this occurs only insofar as the
demand (\ref{30}), which now should be written as 
\begin{equation}
a_{\beta \beta ^{\prime }}a_{\beta ^{\prime }\beta ^{\prime \prime
}}a_{\beta ^{\prime \prime }\beta ^{\prime \prime \prime }}...a_{\beta
^{(n-1)}\beta ^{(n)}}=a_{\beta \beta ^{(n)}},  \label{52}
\end{equation}%
is fulfilled. As has been already said, this can occur only approximately,
and taking it as an exact relation as a result of the application of
Principle Three means neglecting the residual noise.

Eqs.(\ref{46}) for the random phases and (\ref{51}) for the relevant
frequencies have exactly the same structure, so that the mechanism that
solves the former also solves the latter. Thus, to solve Eq.(\ref{51}) we
write%
\begin{equation}
\omega _{\alpha \beta }=\Omega _{\alpha }-\Omega _{\beta }.  \label{53}
\end{equation}%
where the parameters $\Omega _{\lambda }$ are sure numbers to be determined
later (see Eq.(\ref{87b})). It can be easily seen that this form satisfies
Eq.(\ref{51}) identically.

The solution (\ref{53}) is precisely the one arrived at during the
foundations of matrix mechanics \cite{BJH}. There is however an important
difference between that original derivation and the present one. Although in
both cases Eq.(\ref{53}) is used to ensure that each term in a Fourier
development corresponds to the appropriate frequency, in the present case it
is not a formal device required to get the correct Fourier development but a
result of the three principles we have used to construct \textsc{lsed},
i.e., it is a consequence of fundamental postulates and has a physical
meaning over and above its mathematical necessity.

\subsection{The response amplitudes}

Let us now turn to the equations that determine the response functions and
the relevant frequencies. These are the set of equations (\ref{32}), which
in the new notation become 
\begin{subequations}
\begin{equation}
\widetilde{x}_{\alpha \beta }=-\frac{e\widetilde{E}_{\alpha \beta }}{m\omega
_{\alpha \beta }^{2}+im\tau \omega _{\alpha \beta }^{3}+\frac{\widetilde{f}%
_{\alpha \beta }}{\widetilde{x}_{\alpha \beta }a_{\alpha \beta }}},
\label{55a}
\end{equation}%
where $\widetilde{f}_{\alpha \beta }$ is the Fourier component of frequency $%
\omega _{\alpha \beta }$ of the external force. In its turn, Eq.(\ref{25b})
reads now 
\begin{equation}
x_{\alpha }(t)=\sum_{\beta }\frac{-e\widetilde{E}_{\alpha \beta }a_{\alpha
\beta }}{m\omega _{\alpha \beta }^{2}+im\tau \omega _{\alpha \beta }^{3}+%
\frac{\widetilde{f}_{\alpha \beta }}{\widetilde{x}_{\alpha \beta }a_{\alpha
\beta }}}e^{i\omega _{\alpha \beta }t}.  \label{55b}
\end{equation}%
This equation shows that now we have a whole set of solutions, labeled by
the index $\alpha $ that we have been forced to add by following the above
rules. These solutions are obtained by solving the complete set of
simultaneous equations (\ref{55a}), where $\widetilde{f}_{\alpha \beta }$ is
expressed as a function of the response functions themselves and the
relevant frequencies. For example, for an external force expressed as a
power series we would have something like 
\begin{equation}
\widetilde{f}_{\alpha \beta }=\left[ k_{1}\widetilde{x}_{\alpha \beta
}+k_{2}\sum_{\beta ^{\prime }}\widetilde{x}_{\alpha \beta ^{\prime }}%
\widetilde{x}_{\beta ^{\prime }\beta }+k_{3}\sum_{\beta ^{\prime },\beta
^{\prime \prime }}\widetilde{x}_{\alpha \beta ^{\prime }}\widetilde{x}%
_{\beta ^{\prime }\beta ^{\prime \prime }}\widetilde{x}_{\beta ^{\prime
\prime }\beta }+\ldots \right] a_{\alpha \beta }.  \label{55c}
\end{equation}%
Taken as an implicit equation for the response function $\widetilde{x}%
_{\alpha \beta },$ Eq.(\ref{55a}) has in the general case solutions
dominated by its poles. We should expect that at each of the corresponding
frequencies there is a strong response of the mechanical system, and since $%
\tau $ is a very small parameter (normally $\tau \omega _{\alpha \beta }\ll
1,$ so that in the quantum mechanical case the term $im\tau \omega _{\alpha
\beta }^{3}$ coming from the radiation damping is neglected ) this response
is extremely sharp. This suggests to take them as resonances that occur at
the corresponding frequencies, which are just the relevant frequencies.

Eq.(\ref{55b}) shows that the function $x_{\alpha }(t)$ is expressed as a
linear function of the stochastic amplitudes $a_{\alpha \beta };$ the same
is true for $p_{\alpha }(t)=m\overset{.}{\dot{x}}_{\alpha }(t)$ (we are here
again neglecting a small ra\-di\-ative correction). Since the product of any
number of relevant stochastic amplitudes can be expressed as a single
amplitude by applying Eq.(\ref{52}), any dynamic variable, taken as a
function of the $x_{\alpha }(t)$ and $p_{\alpha }(t)$, can in principle be
expressed as a linear function of the amplitudes $a_{\alpha \beta }$ (or, in
exceptional cases, independent of them)$.$ This is the reason for having
named the present theory \emph{linear stochastic electrodynamics}, \textsc{%
lsed}. It is a fundamental and distinctive feature of the theory; it
explains, for example, why all systems described by it (and hence by quantum
mechanics) behave as if they consisted of a set of linear oscillators.

\section{The equations of motion}

The algebraic relationships just obtained between dynamic quantities
strongly suggest to adopt a matrix language to frame the whole theory. Take
Eq.(\ref{55c}) as an example, with the series of terms within the square
brackets contributing to the force term $\widetilde{f}_{\alpha \beta }$.
Each one of these is easily recognizable as a matrix element, e.g. for the
third-order term we have 
\end{subequations}
\begin{equation}
\sum_{\beta ^{\prime },\beta ^{\prime \prime }}\widetilde{x}_{\alpha \beta
^{\prime }}\widetilde{x}_{\beta ^{\prime }\beta ^{\prime \prime }}\widetilde{%
x}_{\beta ^{\prime \prime }\beta }=\left( \widetilde{x}^{3}\right) _{\alpha
\beta }  \label{60}
\end{equation}%
a.s.o., so that also the quantity $\widetilde{f}_{\alpha \beta }/a_{\alpha
\beta }$ can be recognized as a matrix element. Going now back to Eq.(\ref%
{24}) in the new notation, i.e. with the second subindex introduced, we have
for every Fourier component: 
\begin{subequations}
\begin{equation}
-\left( m\omega _{\alpha \beta }^{2}\widetilde{x}_{\alpha \beta }+im\tau
\omega _{\alpha \beta }^{3}\widetilde{x}_{\alpha \beta }+\frac{\widetilde{f}%
_{\alpha \beta }}{a_{\alpha \beta }}\right) e^{i\omega _{\alpha \beta }t}=e%
\widetilde{E}_{\alpha \beta }e^{i\omega _{\alpha \beta }t},  \label{61a}
\end{equation}%
which is itself an equation relating matrix elements. Associating the
elementary oscillator $e^{i\omega _{\alpha \beta }t}$ to each one of the
matrix elements we can rewrite this as a dynamic equation: 
\begin{equation}
m\frac{d^{2}\widetilde{x}_{\alpha \beta }(t)}{dt^{2}}=\widetilde{f}_{\alpha
\beta }(t)+m\tau \frac{d^{3}\widetilde{x}_{\alpha \beta }(t)}{dt^{3}}+e%
\widetilde{E}_{\alpha \beta }(t).  \label{61b}
\end{equation}%
with 
\begin{equation}
\widetilde{x}_{\alpha \beta }(t)=\widetilde{x}_{\alpha \beta }e^{i\omega
_{\alpha \beta }t},\quad \widetilde{E}_{\alpha \beta }(t)=\widetilde{E}%
_{\alpha \beta }e^{i\omega _{\alpha \beta }t},\quad \widetilde{f}_{\alpha
\beta }(t)=\frac{\widetilde{f}_{\alpha \beta }}{a_{\alpha \beta }}e^{i\omega
_{\alpha \beta }t}.  \label{61c}
\end{equation}%
In closed matrix notation, Eq.(\ref{61b}) reads 
\begin{equation}
m\frac{d^{2}\widehat{x}}{dt^{2}}=\widehat{f}+m\tau \frac{d^{3}\widehat{x}}{%
dt^{3}}+e\widehat{E}.  \label{61d}
\end{equation}%
The field operator written in full is 
\begin{equation}
\widehat{E}\left( \omega \right) =iN\sqrt{\mathcal{E}}\left( \widehat{a}%
e^{i\omega t}-\widehat{a}^{\dagger }e^{-i\omega t}\right) ,  \label{61db}
\end{equation}%
with 
\begin{equation}
\widehat{a}=-\frac{i\widetilde{E}^{(+)}}{N\sqrt{\mathcal{E}}}a,\quad 
\widehat{a}^{\dagger }=\frac{i\widetilde{E}^{(-)}}{N\sqrt{\mathcal{E}}}%
a^{\ast },  \label{61dc}
\end{equation}%
$\mathcal{E}=\frac{1}{2}\hbar \omega $ and $N$ a suitable normalizing
factor, as explained below ($\widetilde{E}^{(\pm )}$ are now matrices). In
writing these equations we have separated positive and negative frequencies
for clarity (in Eq.(\ref{61db}) $\omega >0$).

Eq.(\ref{61b}) is the law of motion for the mechanical subsystem in the
quantum regime according to \textsc{lsed}; it agrees with the corresponding
equation of nonrelativistic quantum electrodynamics \cite{Cohen}. If now we
neglect the field and the radiation reaction terms to get a purely
mechanical description, which can be done since owing to the principles
under which the theory has been constructed, the field has already played
its central role in stabilizing the atomic subsystem and driving it to the
quantum regime, we get the usual Heisenberg equations of motion of
nonrelativistic quantum mechanics, 
\begin{equation}
\frac{d\widehat{p}}{dt}=\widehat{f},  \label{61e}
\end{equation}%
\begin{equation}
\widehat{p}=m\frac{d\widehat{x}}{dt}.  \label{61f}
\end{equation}%
The neglected terms, when reintroduced, lead to the (well-known) radiative
corrections arising both from the background field and radiation reaction.

\subsection{Completing the description}

Establishing the full equivalence between quantum mechanics and \textsc{lsed}
still requires some additional results, to which we now pay attention. We
start by considering the Poisson brackets of dynamical variables in the 
\textsc{lsed} description. Originally the configuration and momentum
coordinates are the $x,$ $p$ of the particle (or particles) and the $%
q_{\alpha \beta },$ $p_{\alpha \beta }$ of the field. However, once the
quantum regime is established, the particle variables $x_{\alpha },$ $%
p_{\alpha }$ corresponding to a stationary state are not any more
independent, as they have become functions of the field variables $q_{\alpha
\beta },$ $p_{\alpha \beta }$ or, equivalently, of the random amplitudes $%
a_{\alpha \beta },$ $a_{\alpha \beta }^{\ast },$ as follows from Eq.(\ref%
{22e2}), which in full notation reads 
\end{subequations}
\begin{equation}
p_{\alpha \beta }=\sqrt{\frac{\mathcal{E}_{\alpha \beta }}{2}}\left(
a_{\alpha \beta }+a_{\alpha \beta }^{\ast }\right) ,\quad i\omega _{\alpha
\beta }q_{\alpha \beta }=\sqrt{\frac{\mathcal{E}_{\alpha \beta }}{2}}\left(
a_{\alpha \beta }-a_{\alpha \beta }^{\ast }\right) .  \label{65}
\end{equation}%
with $\mathcal{E}_{\alpha \beta }=\frac{1}{2}\hbar \omega _{\alpha \beta }$.
This means that the Poisson bracket of the couple of dynamical variables $%
A_{\alpha },B_{\alpha }$ should be written as%
\begin{equation*}
\left[ A,B\right] _{PB}=\sum_{\lambda }\left[ \frac{\partial A}{\partial
q_{\alpha \lambda }}\frac{\partial B}{\partial p_{\alpha \lambda }}-\frac{%
\partial A}{\partial p_{\alpha \lambda }}\frac{\partial B}{\partial
q_{\alpha \lambda }}\right] 
\end{equation*}%
\begin{equation*}
=\sum_{\lambda }-\frac{i\omega _{\alpha \lambda }}{2\mathcal{E}_{\alpha
\lambda }}\left[ \frac{\partial A}{\partial a_{\alpha \lambda }}\frac{%
\partial B}{\partial a_{\alpha \lambda }^{\ast }}-\frac{\partial A}{\partial
a_{\alpha \lambda }^{\ast }}\frac{\partial B}{\partial a_{\alpha \lambda }}%
\right] 
\end{equation*}%
\begin{subequations}
\begin{equation}
=\frac{1}{i\hbar }\sum_{\lambda }\left[ \frac{\partial A}{\partial a_{\alpha
\lambda }}\frac{\partial B}{\partial a_{\alpha \lambda }^{\ast }}-\frac{%
\partial A}{\partial a_{\alpha \lambda }^{\ast }}\frac{\partial B}{\partial
a_{\alpha \lambda }}\right] ,  \label{66a}
\end{equation}%
where both variables $A$ and $B$ are understood to refer to the state $%
\alpha .$ The quantity appearing in the last equality within brackets was
introduced in a similar context in \cite{dlpc86} with the name \emph{%
Poissonian}, and in a different context under the name of \emph{commutator}
in \cite{ES83b}; denoting it with $\left\langle A;B\right\rangle $ (the
index $\alpha $ is implied as above) we get 
\begin{equation}
\left[ A,B\right] _{PB}=\frac{1}{i\hbar }\left\langle A;B\right\rangle .
\label{66b}
\end{equation}%
It is easy to generalize Eq.(\ref{66a}) to include nondiagonal elements, by
writing 
\end{subequations}
\begin{equation*}
\left\langle A;B\right\rangle _{\alpha \beta }=\sum_{\lambda }\left[ \frac{%
\partial A}{\partial a_{\alpha \lambda }}\frac{\partial B}{\partial a_{\beta
\lambda }^{\ast }}-\frac{\partial A}{\partial a_{\beta \lambda }^{\ast }}%
\frac{\partial B}{\partial a_{\alpha \lambda }}\right] a_{\alpha \beta }
\end{equation*}%
\begin{equation}
=\sum_{\lambda }\left[ \frac{\partial A}{\partial a_{\alpha \lambda }}\frac{%
\partial B}{\partial a_{\lambda \beta }}-\frac{\partial B}{\partial
a_{\alpha \lambda }}\frac{\partial B}{\partial a_{\lambda \alpha }}\right]
a_{\alpha \beta }.  \label{67}
\end{equation}%
In writing this equation we have taken into account that from Eq.(\ref{50})
it follows that $a_{\alpha \beta }^{\ast }=a_{\beta \alpha };$ similarly, $%
\omega _{\alpha \beta }=-\omega _{\beta \alpha }.$ This result can be recast
immediately in terms of the matrix elements of the variables $A$ and $B$.
Indeed, by writing 
\begin{equation}
A=\sum \widetilde{A}_{\alpha \lambda }a_{\alpha \lambda }e^{i\omega _{\alpha
\lambda }t},\quad B=\sum \widetilde{B}_{\alpha \lambda }a_{\alpha \lambda
}e^{i\omega _{\alpha \lambda }t},  \label{68}
\end{equation}%
we get successively%
\begin{equation}
\left\langle A;B\right\rangle _{\alpha \beta }=\sum_{\lambda }\left[ 
\widetilde{A}_{\alpha \lambda }\widetilde{B}_{\lambda \beta }-\widetilde{B}%
_{\alpha \lambda }\widetilde{A}_{\lambda \beta }\right] a_{\alpha \beta
}=\left( \widetilde{A}\widetilde{B}-\widetilde{B}\widetilde{A}\right)
_{\alpha \beta }a_{\alpha \beta }=\left[ \widehat{A},\widehat{B}\right]
_{\alpha \beta }a_{\alpha \beta }.  \label{69}
\end{equation}%
In the last expression we have introduced the commutator of the matrices $%
\widehat{A}$ and $\widehat{B}$ with matrix elements%
\begin{equation}
\left[ \widehat{A},\widehat{B}\right] _{\alpha \beta }=\sum_{\lambda }\left[ 
\widetilde{A}_{\alpha \lambda }\widetilde{B}_{\lambda \beta }-\widetilde{B}%
_{\alpha \lambda }\widetilde{A}_{\lambda \beta }\right] .  \label{70}
\end{equation}%
We have thus found the correspondences%
\begin{equation}
\left[ A,B\right] _{PB}\leftrightarrow \frac{1}{i\hbar }\left\langle
A;B\right\rangle \leftrightarrow \frac{1}{i\hbar }\left[ \widehat{A},%
\widehat{B}\right] .  \label{71}
\end{equation}%
Two important applications are the following: 
\begin{equation}
\text{The identity }\left[ x,p\right] _{PB}=1\text{ leads to the fundamental
commutator }\left[ \widehat{x},\widehat{p}\right] =i\hbar \widehat{1}.
\label{72}
\end{equation}%
\begin{equation}
\text{The eq. of motion }\frac{dA}{dt}\hspace{-0.06cm}=\hspace{-0.06cm}\left[
A,H\right] _{PB}\text{ leads to the Heisenberg eq. }i\hbar \frac{d\widehat{A}%
}{dt}\hspace{-0.06cm}=\hspace{-0.06cm}\left[ \widehat{A},\widehat{H}\right] .
\label{73}
\end{equation}%
It must be noted that whereas the classical equation $\left[ x,p\right]
_{PB}=1$ is an identity, the corresponding commutator $\left[ \widehat{x},%
\widehat{p}\right] =i\hbar $ is a \emph{derived} equation that holds only in
the quantum regime. It has a dynamic meaning, precisely because it implies
that the mechanical system has already reached the quantum regime, in which
the mechanical variables are driven by the field variables; in other words,
it is a physical law. From Eq.(\ref{65}) it is easy to see that with the
definition given above, Eq.(\ref{72}) holds also for the field variables
(because they describe quantum oscillators), so that we recover the usual
rule 
\begin{equation}
\left[ \widehat{a},\widehat{a}^{\dagger }\right] =-\frac{i\omega }{2\mathcal{%
E}}\left[ \widehat{q},\widehat{p}\right] =\frac{1}{i\hbar }\left[ \widehat{q}%
,\widehat{p}\right] =\widehat{1}.  \label{74}
\end{equation}%
Another related point that merits some attention is the following. The
matrix elements of any dynamical variable are given by equations as (\ref%
{55a}) or (\ref{61a}), which contain no arbitrary elements in principle. In
quantum mechanics, however, as there is no explicit reference to the vacuum
field components $\widetilde{E}$, the scale of the matrix elements is lost.
This problem is solved by normalizing the state vectors to unity (and so $%
\left\langle \alpha \right\vert \left. \beta \right\rangle =\delta _{\alpha
\beta }$). In the present theory, such scale can be introduced by means of
Eq.(\ref{74}), which fixes the normalization factor $N$ that was left
undetermined in Eq.(\ref{61db}).

It remains still to determine the meaning of the parameter $\Omega _{\lambda
}$ introduced in Eq.(\ref{53}). To achieve this we combine this equation
with (\ref{61c}) to write%
\begin{equation*}
\widetilde{\overset{.}{x}}_{\alpha \beta }=i\omega _{\alpha \beta }%
\widetilde{x}_{\alpha \beta }=i\left( \Omega _{\alpha }-\Omega _{\beta
}\right) \widetilde{x}_{\alpha \beta } 
\end{equation*}%
\begin{equation}
=i\sum_{\lambda }\left( \Omega _{\alpha }\delta _{\alpha \lambda }\widetilde{%
x}_{\lambda \beta }-\widetilde{x}_{\alpha \lambda }\Omega _{\beta }\delta
_{\lambda \beta }\right) .  \label{85}
\end{equation}%
On the other hand, from the equation of motion (\ref{73}) it follows that%
\begin{equation}
i\hbar \widetilde{\overset{.}{x}}_{\alpha \beta }=\sum_{\lambda }\left( 
\widetilde{x}_{\alpha \lambda }\widetilde{H}_{\lambda \beta }-\widetilde{H}%
_{\alpha \lambda }\widetilde{x}_{\lambda \beta }\right) .  \label{86}
\end{equation}%
A comparison gives 
\begin{subequations}
\begin{equation}
\widetilde{H}_{\alpha \beta }=\hbar \Omega _{\alpha }\delta _{\alpha \beta
}+c\widetilde{x}_{\alpha \beta }+d\delta _{\alpha \beta },  \label{87a}
\end{equation}%
with $c$ and $d$ arbitrary. By using $\widetilde{\overset{.}{p}}_{\alpha
\beta }=i\omega _{\alpha \beta }\widetilde{p}_{\alpha \beta }$ and applying
a similar procedure we conclude that necessarily $c=0$. The constant $d$
simply shifts the overall reference level of $H$, and can therefore be
dropped. Therefore, the matrix representing the Hamiltonian in state $\alpha 
$ is diagonal and has sure values, 
\begin{equation}
H_{\alpha }=\sum_{\lambda }H_{\alpha \lambda }a_{\alpha \lambda }=\hbar
\Omega _{\alpha }a_{\alpha \alpha }=\hbar \Omega _{\alpha }\equiv E_{\alpha
},  \label{87b}
\end{equation}%
as was to be expected, since the stationary states were defined from the
very start (Eq. \ref{wa}) as those for which energy equilibrium had been
reached. In terms of the energy $E_{\alpha }$ of state $\alpha $, Eq.(\ref%
{53}) becomes Bohr's rule for the transition frequencies, 
\begin{equation}
\hbar \omega _{\alpha \beta }=E_{\alpha }-E_{\beta }.  \label{87c}
\end{equation}%
One can thus identify the relevant frequencies with the corresponding
quantum transition frequencies. Since these coincide with the spectroscopic
frequencies, Eq.(\ref{87c}) corresponds to the old Ritz principle, stating
that each spectroscopic frequency can be written as the difference of two
terms. In the present theory, however (as in quantum theory) this is a
prediction. It is important to observe that Principle Three, by assigning a
sure value to $\omega _{\alpha \beta }$, concurrently assigns sure values to
the energy, which correspond to the eigenvalues of the Hamiltonian, as
stated in Eq.(\ref{87b}) (although in a different language). Principle Three
can therefore be considered as the quantization principle, a point on which
we elaborate below. In this form we have verified that \textsc{lsed} is
formally equivalent to (nonrelativistic) \textsc{qed} and to quantum
mechanics in the radiationless approximation.

Recently, Cole and Zou \cite{ColZou03} obtained a series of appealing
numerical results for the ground state of the hydrogen atom directly from
the principles of \textsc{sed}, having a strong resemblance with the
corresponding predictions of quantum mechanics. Their computations coincide
in spirit with the present theory, since both approaches are based on the
principles of \textsc{sed} but are stripped from the old methodological
assumptions, so neither perturbative nor Fokker-Planck methods are being
used. Thus the present work gives theoretical backing to their results,
while at the same time it is at least in part underpinned by their numerical
experiments.

Still the conceptual differences between \textsc{lsed} and quantum mechanics
are momentous. A brief discussion of these matters is given in the
Discussion section at the end. For the time being let us just briefly remark
that from the present point of view, quantum mechanics furnishes an
approximate, time-asymptotic statistical description of the mechanical
(atomic) part of the system under study, valid once the quantum regime has
set in. The passage to \textsc{qed} improves the description by adding part
of the lost noise and by leading to matter and field quantization, which
calls for new phenomena. But even then, the description continues to be
approximate, statistical and time asymptotic. Only a return to the initial,
complete description could lead to a qualitative improvement of the account.
This is a task that pertains to the future.

\section{Detailed energy balance}

Now we come back to Eq.(\ref{wb}) 
\end{subequations}
\begin{equation*}
m\tau \left\langle \overset{\cdot \cdot }{\mathbf{x}}^{2}\right\rangle
=e\left\langle \overset{\cdot }{\mathbf{x}}\cdot \mathbf{E}\right\rangle 
\end{equation*}%
describing the average power balance. Our first undertaking will be to give
to this equation a more finished form. The quantum mechanical solutions
follow from Eq.(\ref{61e}), but they must be amended using Eq.(\ref{61d}) to
take into account the radiative corrections and other phenomena. Treating
the corrections as a perturbation, the solution will read (once more in one
dimension) $x=x_{0}+x_{1},$ where $x_{0}$ represents the unperturbed
solution and $x_{1}$ its correction. The average absorbed and radiated power
are now $m\tau \left\langle \overset{\cdot \cdot }{x}_{0}^{2}\right\rangle $
and $e\left\langle \left( \overset{.}{x}_{0}+\overset{.}{x}_{1}\right)
E\right\rangle ,$ respectively, where we have neglected the small
corrections to the radiation.

Let us now assume that $x_{0}$ is proportional to $e^{r},$ whereas $x_{1}$
is proportional to $e^{s}$. Here $e$ stands for the coupling constant to the
radiation field, and we should carefully distinguish it from the possible
appearance of the charge in the external force, which in the present context
appears as merely incidental. Thus, the charge $e$ that appears here is
foreign to quantum mechanics, as it is linked to the background field $E$.
Further, since $m\tau =2e^{2}/3c^{3},$ the power radiated is proportional to 
$e^{2r+2},$ whereas the power absorbed due to $ex_{0}$ is proportional to $%
e^{r+1}.$ For these two quantities to be equal, we must have $r=-1.$ But
this is contrary to quantum mechanics, since with the normalization used
there, one should have $r=0$, as has just been argued$.$ Thus$,$ the term $%
e\left\langle \overset{.}{x}_{0}E\right\rangle $ cannot contribute to the
energy absorption. However the term proportional to $x_{1}$ requires that $%
2r+2=s+1,$ or $s=2r+1.$ Putting here $r=0$ we get $s=1.$ As will be shown,
this corresponds exactly to the correction $x_{1}$ (see Eq.(\ref{101})), so
we conclude that the equation for the energy balance, omitting the spurious
term, reads%
\begin{equation}
m\tau \left\langle \overset{\cdot \cdot }{\mathbf{x}}_{0}^{2}\right\rangle
=e\left\langle \overset{\cdot }{\mathbf{x}_{1}}\cdot \mathbf{E}\right\rangle
.  \label{100}
\end{equation}%
A more formal argument to arrive at this equation goes as follows. The
quantity $\overset{.}{x}_{0}E$ belongs to \textsc{qed}, where the vacuum
field is a natural element of the theory and the term accounts for the
effect of the radiative correction. However, $\overset{.}{x}_{0}$ and $E$
belong to different Hilbert spaces, so that $\left\langle \overset{.}{x}%
_{0}E\right\rangle $ is proportional to $\left\langle 0\right| E\left|
0\right\rangle =0,$ and the contribution effectively cancels out.

Incidentally, we note that Eq.(\ref{100}) is independent not just from the
charge, that is, from the strength of the particle's coupling to the vacuum
field (we stress that $x_{1}=\delta x$ in Eq.(\ref{101}) is proportional to $%
e$), but also from the mass of the particle. This strongly suggests a
principle of universality, according to which the variance of the
acceleration is largely independent from the specific details of the
particle and perhaps, of the interaction. This principle has already been
advocated from different considerations within \textsc{sed} \cite{ES2}; and
to the extent that it holds, \textsc{sed} would be but a particular version
of a more general theory, in which different kinds of vacuum field may
participate (of course, all of them with the same average energy per mode
and hence in equilibrium among them). In its turn, this points to the
possibility that a more general theory could be formulated not in terms of
random vacua, but of a fluctuating metric, which would be a truly universal
theory \cite{Dice}$.$

The next step is the determination of $x_{1},$ which constitutes a
conventional problem readily solved using perturbation theory. The result to
first order is \cite{Dice} (we return to our previous notation, so that $%
x_{0}$ is denoted by $x,$ and instead of $x_{1}$ we write $\delta x$)%
\begin{equation}
\delta x_{\alpha }\left( t\right) =-\frac{2e}{\hbar }\sum_{\beta }\left\vert 
\widetilde{x}_{\alpha \beta }\right\vert ^{2}\int_{0}^{\infty }E\left(
t-s\right) \sin \omega _{\alpha \beta }s\,ds.  \label{101}
\end{equation}%
We now take into account that \cite{Dice} 
\begin{subequations}
\begin{equation}
\left\langle E_{i}(\mathbf{x},t)E_{j}(\mathbf{x},t^{\prime })\right\rangle
=\delta _{ij}\int_{0}^{\infty }S\left( \omega \right) \cos \omega \left(
t-t^{\prime }\right) d\omega ,  \label{102a}
\end{equation}%
\begin{equation}
S\left( \omega \right) =\frac{4\pi }{3}\rho \left( \omega \right) ,
\label{102b}
\end{equation}%
where $S\left( \omega \right) $ is the power spectrum of the vacuum field
and $\rho \left( \omega \right) $ is its (energy) spectral density.
Inserting these results into the expression for the average power absorbed
and performing the integration, we get 
\end{subequations}
\begin{equation}
e\left\langle \delta \overset{.}{x}E\right\rangle _{\alpha }=-\frac{4\pi
^{2}e^{2}}{3\hbar }\sum_{\beta }\omega _{\alpha \beta }\rho \left( \omega
_{\alpha \beta }\right) \left\vert \widetilde{x}_{\alpha \beta }\right\vert
^{2}.  \label{103}
\end{equation}%
In its turn, in the quantum regime the average power radiated is $\left(
2e^{2}/3c^{3}\right) \left\langle \overset{..}{x}^{2}\right\rangle $ $\
=\left( 2e^{2}/3c^{3}\right) \omega _{\alpha \beta }^{4}\left\vert 
\widetilde{x}_{\alpha \beta }\right\vert ^{2}.$ Thus Eq.(\ref{100})
transforms into%
\begin{equation}
\sum_{\beta }\frac{2e^{2}}{3c^{3}}\left[ -\left\vert \omega _{\alpha \beta
}\right\vert ^{3}+\frac{2\pi ^{2}c^{3}}{\hbar }\rho \left( \omega _{\alpha
\beta }\right) \right] \left\vert \omega _{\alpha \beta }\right\vert
\left\vert \widetilde{x}_{\alpha \beta }\right\vert ^{2}=0.  \label{104}
\end{equation}%
In writing this equation we assumed that $\omega _{\alpha \beta }$ is
negative, as is the case for the ground state. We assumed also that the
frequencies $\omega _{\alpha \beta }$ are nondegenerate. This equation is
satisfied irrespective of the coefficients $\widetilde{x}_{\alpha \beta },$
which means without regard to the specific system under study, if the
expression within brackets vanishes for every $\omega _{\alpha \beta },$ or
if%
\begin{equation}
\rho \left( \omega \right) =\rho _{0}\left( \omega \right) \equiv \frac{%
\hbar \omega ^{3}}{2\pi ^{2}c^{3}}\quad \left( \omega >0\right) .
\label{105}
\end{equation}%
This is just the spectral density of a vacuum with average energy per mode $%
\hbar \omega /2,$ so it corresponds to that of the vacuum field of \textsc{%
sed} (and \textsc{qed}). This result means that indeed the balance equation (%
\ref{100}) is satisfied by each frequency separately (whether or not there
is degeneracy), or that \emph{detailed} energy balance holds for any bounded
system described by \textsc{lsed}. Alternatively, the argument can be seen
as a derivation of the zero-point spectrum from the requirement of detailed
balance. The result is important, not only because it shows the internal
consistency of the theory, but also on the account that it stands in sharp
contrast with the corresponding classical result for a general system with
harmonics, where detailed balance holds only for the Rayleigh-Jeans spectrum 
$\rho \left( \omega \right) \sim \omega ^{2}$ and, perhaps worse, only for a
Laplacian distribution of energy (Maxwell-Boltzmann statistics) \cite{vVH}.

\section{Some generalizations. Planck's distribution}

With the purpose of providing a more general perspective of the theory we
give some generalizations of the above results, without however entering
into their detailed derivations, which can be found elsewhere \cite{Dice}.
Let us first consider a dynamic variable $\xi $ that represents an integral
of motion of the unperturbed system. It is possible to demonstrate that to
first order in perturbation theory the equilibrium condition for this
variable reads%
\begin{equation}
\sum_{\beta }\left[ -\frac{\hbar }{2\pi ^{2}c^{3}}\omega _{\alpha \beta
}^{3}+\rho \left( \omega _{\alpha \beta }\right) \right] \left( \xi _{\alpha
}-\xi _{\beta }\right) \left| \widetilde{x}_{\alpha \beta }\right| ^{2}=0.
\label{110}
\end{equation}%
For $\xi =H$ this result reduces to Eq.(\ref{104}), as it should. For any
other integral of motion for which $\xi _{\alpha }\neq \xi _{\beta }$ Eq.(%
\ref{110}) leads to the same equilibrium spectral energy density $\rho
_{0}(\omega )$, as should be expected in advance.

Let us now extend the result (\ref{104}) to cover the case of excited states
and a more general external random electromagnetic field, to study the
equilibrium conditions. We write the spectral energy density of the field in
the form $\rho =\rho _{e}+\rho _{0},$ where $\rho _{0}$ is given by Eq.(\ref%
{105}) and $\rho _{e}$ represents the spectral density of the field above
the zero-point. We can write now%
\begin{equation}
\left\langle \frac{dH}{dt}\right\rangle =-\frac{4\pi ^{2}e^{2}}{3\hbar }%
\sum_{\beta }\omega _{\alpha \beta }\left[ \rho +\text{sign}\left( \omega
_{\alpha \beta }\right) \rho _{0}\right] \left| \widetilde{x}_{\alpha \beta
}\right| ^{2}.  \label{111}
\end{equation}%
We now separate positive and negative frequencies, adding a superindex $\pm $
to $\widetilde{x}_{\alpha \beta }$ to keep track of this sign,%
\begin{equation}
\left\langle \frac{dH}{dt}\right\rangle =-\frac{4\pi ^{2}e^{2}}{3\hbar }%
\sum_{\beta }-\omega _{\alpha \beta }\left[ \left( \rho -\rho _{0}\right)
\left| \widetilde{x}_{\alpha \beta }^{\left( -\right) }\right| ^{2}+\left(
\rho +\rho _{0}\right) \left| \widetilde{x}_{\alpha \beta }^{\left( +\right)
}\right| ^{2}\right] ,  \label{112}
\end{equation}%
and recast the result into the form 
\begin{subequations}
\begin{equation}
\left\langle \frac{dH}{dt}\right\rangle =W_{\text{ab}}-W_{\text{em}},
\label{115a}
\end{equation}%
where 
\begin{equation}
W_{\text{ab}}=\frac{4\pi ^{2}e^{2}}{3\hbar }\sum_{\omega _{\alpha \beta
}<0}\left| \omega _{\alpha \beta }\right| \left( \rho -\rho _{0}\right)
\left| \widetilde{x}_{\alpha \beta }^{\left( -\right) }\right| ^{2},
\label{115b}
\end{equation}%
\begin{equation}
W_{\text{em}}=\frac{4\pi ^{2}e^{2}}{3\hbar }\sum_{\omega _{\alpha \beta
}>0}\omega _{\alpha \beta }\left( \rho +\rho _{0}\right) \left| \widetilde{x}%
_{\alpha \beta }^{\left( +\right) }\right| ^{2}  \label{115c}
\end{equation}%
are the contributions of the absorptions and emissions to the energy change,
respectively. Eq.(\ref{115b}) clearly shows that for absorptions to occur,
necessarily $\rho >\rho _{0},$ i.e., $\rho _{e}$ must be present. Thus,
there are no `spontaneous absorptions', $W_{\text{ab}}^{\text{spont}}=0$ in
the present theory, just as happens in \textsc{qed} and in nature, of
course. This behavior is due to the fact, clearly shown in Eq.(\ref{115b}),
that the ground state is just that supported by the vacuum field; to get
into a higher state the atomic system should be immersed in a field with $%
\rho >\rho _{0}.$ It is also interesting to have a closer look at Eq.(\ref%
{115c}) for the probability of an emission to take place. As follows from
Eq.(\ref{104}) the contribution containing $\rho _{0}$ comes from the
effects of radiation reaction, whereas the term that involves the whole
spectral density $\rho $ is due to the fluctuating motions. For a pure
vacuum $\rho =\rho _{0}$ both contributions become alike and contribute with
equal amounts to the emissions, whereas they exactly cancel out for
absorptions. Of course this latter result is but another form to express the
fact that the system has reached the quantum regime with the vacuum \cite%
{note1}. This is an important point because \textsc{sed} (and presumably 
\textsc{lsed}) has been charged of being a semiclassical theory and thus
necessarily predicting spontaneous absorptions \cite{Milonni}. In fact all
absorptions are induced with probability 
\begin{equation}
W_{\text{ab}}^{\text{ind}}=\frac{4\pi ^{2}e^{2}}{3\hbar }\sum_{\omega
_{\alpha \beta }<0}\left| \omega _{\alpha \beta }\right| \rho _{e}\left| 
\widetilde{x}_{\alpha \beta }^{\left( -\right) }\right| ^{2}.  \label{120a}
\end{equation}%
On the other hand, writing $\rho +\rho _{0}=\rho _{e}+2\rho _{0}$ in Eq.(\ref%
{115c}) we obtain induced and spontaneous emissions, the latter being due
solely to the action of the vacuum field,%
\begin{equation}
W_{\text{em}}^{\text{ind}}=\frac{4\pi ^{2}e^{2}}{3\hbar }\sum_{\omega
_{\alpha \beta }>0}\omega _{\alpha \beta }\rho _{e}\left| \widetilde{x}%
_{\alpha \beta }^{\left( +\right) }\right| ^{2},  \label{120b}
\end{equation}%
\begin{equation}
W_{\text{em}}^{\text{spont}}=\frac{8\pi ^{2}e^{2}}{3\hbar }\sum_{\omega
_{\alpha \beta }>0}\omega _{\alpha \beta }\rho _{0}\left| \widetilde{x}%
_{\alpha \beta }^{\left( +\right) }\right| ^{2}.  \label{120c}
\end{equation}%
From these results (or their generalization to any other integral of motion)
it is easy to obtain the Planck distribution as the equilibrium solution for 
$\rho _{e}$ by following the well known statistical method introduced by
Einstein. For this purpose, let us consider a system with only two active
levels, so that a single frequency, which we call $\omega _{\alpha \beta },$
is relevant. To support the state of thermodynamic equilibrium the system
must be embedded in an appropriate field that allows for upward and downward
transitions to occur at the same constant rate. Since the system is in
thermodynamic equilibrium, the populations of the levels $\alpha $ and $%
\beta $ should be proportional to $e^{-\beta E_{\alpha }}$ and $e^{-\beta
E_{\beta }},$ respectively, with $\beta =1/\left( k_{B}T\right) $ (we are
neglecting the possibility of degeneracies, as they would add nothing but
complications to the argument). Thus from the equilibrium condition $W_{%
\text{ab}}=W_{\text{em}}$ applied to Eqs.(\ref{120a})-(\ref{120c}), we get 
\end{subequations}
\begin{equation}
e^{-\beta E_{\alpha }}\rho _{e}\left| \widetilde{x}_{\alpha \beta }^{\left(
-\right) }\right| ^{2}=e^{-\beta E_{\beta }}\left( \rho _{e}+2\rho
_{0}\right) \left| \widetilde{x}_{\alpha \beta }^{\left( +\right) }\right|
^{2},  \label{121}
\end{equation}%
and since $\left| \widetilde{x}_{\alpha \beta }^{\left( -\right) }\right|
^{2}=\left| \widetilde{x}_{\beta \alpha }^{\left( +\right) }\right|
^{2}=\left| \widetilde{x}_{\alpha \beta }\right| ^{2},$ this gives for the
equilibrium condition%
\begin{equation}
e^{-\beta E_{\alpha }}\rho _{e}=e^{-\beta E_{\beta }}\left( \rho _{e}+2\rho
_{0}\right) ,  \label{122}
\end{equation}%
which in its turn leads to the blackbody distribution (with zero-point
field, of course)%
\begin{equation}
\rho =\rho _{0}+\rho _{e}=\rho _{0}\cosh \left( \tfrac{1}{2}\hbar \omega
\beta \right) .  \label{123}
\end{equation}%
A nice point of this derivation is that it clearly exhibits the Planck
distribution as a universal result, independent of the nature and specific
properties of the material system, since the only elements in Eq.(\ref{121})
referring to such system, the $\left| \widetilde{x}_{\alpha \beta }\right|
^{2},$ cancel out to lead to Eq.(\ref{122}). Also, from Eq.(\ref{112}) we
observe that in the present theory this result comes from the quantum
properties of matter, not those of the field. This is an interesting point,
since it is traditional to consider the Planck distribution as the first
known illustration of the quantum properties of the radiation field. A
similar argument is known in other instances, as is the case with the
photoelectric effect. This effect was explained by Einstein as arising from
the quantum properties of the radiation field, and has been since then taken
as such. However there have been solid arguments \cite{Lamb} to show that
this effect can equally well be interpreted as arising from the quantum
properties of matter. Since as we have seen (and is well known),
quantization of matter and of the radiation field imply one another so they
go together, the coexistence of both possible points of view is
understandable.

Of course the equivalent calculation is well known in quantum theory. What
we are trying to stress with the present reckoning is that \textsc{lsed}
furnishes the results of quantum mechanics and (nonrelativistic) \textsc{qed}
in a quite direct and transparent way. It is possible to go even further
with the calculation of the radiative corrections (the Lamb shift and others 
\cite{Dice}, \cite{cavity}), but for the purposes of illustration the above
examples should suffice. We thus conclude that the principles used to
construct \textsc{lsed} are sufficient to transform an apparently classical
theory into a sound quantum theory. The reason of this seemingly miraculous
transformation is twofold. Firstly, the theory contains a crucial
ingredient, the vacuum field, foreign to classical physics and with
statistical properties specified by $\hbar $, which makes the theory stricto
sensu a nonclassical one from the start ---or a quantum one, as
substantiated by the end results. Secondly, the set of principles used to
develop the theory, particularly Principle Three, is strong enough as to
select a class of (approximate) solutions to the equations of motion that
corresponds just to the quantum behavior. These reasons explain our proviso 
\emph{apparently} classical, used to stress the fact that if it were a plain
classical theory, it would be impossible to derive quantum results from it.

\section{Discussion\label{disc}}

We have arrived at quantum mechanics from a fresh point of view. Even if for
utilitarian purposes the present derivation may seem to be of limited
interest, on the conceptual level it has the benefit of providing a new and
valuable perspective to the foundations of quantum theory. For example, the
theory furnishes a physical explanation on the origin of quantization as due
to the selection of allowed solutions as the robust ones against
fluctuations, \textit{i.e.}, the ones that are immune to small changes in
the particular realization of the vacuum field, once the system has entered
the quantum regime. Further, the new picture elucidates the mechanism
leading to atomic stability, a problem that has puzzled physicists for
nearly a century (we elaborate further on these and related points below and
in the Appendix). But in addition, there appear some points where the
predictions of the theory may permit someday to explore as yet unknown
territories. For the time being let us make a brief tour from the
perspective afforded by the present theory onto some of those traits of
quantum mechanics that have been the subject of endless discussions and
controversies on the matter.

In the usual perspective, quantum mechanics constitutes both the point of
departure and the final reference for our inquiries about the meaning of the
theory itself. Its conceptual problems must therefore be looked at from the
inside, which creates a kind of circular reasoning leading almost nowhere,
as is amply testified by endless discussions on such subjects. Since the
point of departure for \textsc{lsed} is a wider physical theory, it offers a
qualitatively different opportunity. This fact allows in principle to answer
such conceptual questions with a fresh and deeper understanding from an
`external' perspective, without the need to resort to philosophical or
ideological preconceptions. We now attempt to exploit these possibilities to
address, albeit very succinctly and in a schematic fashion, some of the most
abiding quantum questions.

\subsection{Atomic stability}

One point of the proposed theory that surely catches the reader's attention
is the one related to atomic stability. In usual quantum theory the
stationary atomic levels are well predicted by the equations, but the
physical reason for their stability remains undisclosed. In \textsc{lsed}
they appear as those states which comply with the requirement of belonging
to the quantum regime, that is, those for which the rate of radiated and
absorbed energy is the same in the mean. That the levels belong normally to
a discrete spectrum comes from the fact that only for certain orbital
motions such equilibrium can be attained, as follows clearly from Eq.(\ref%
{100}). In quantum mechanics such an explanation is impossible, since there
is no field from which to absorb energy, and thus the point remains as a
mystery, and can find only a formal answer.

\subsection{Energy eigenvalues and the origin of quantization\label{stab}}

A related point is that of the energy (and other) eigenvalues. That in an
essentially stochastic theory the dynamical variables may attain sure
values, seems to be a contradiction, or at least a very obscure property.
The answer that can be derived from the present development goes as follows.
In the first place, the quantum description is approximate, since Principle
Three cannot be satisfied exactly by natural systems; and it is just this
principle that leads to the existence of eigenvalues. Thus, nature is
noisier than what the present theoretical description asserts (which,
indeed, must be corrected by adding at least the field that gives rise to
the radiative corrections, a very special kind of noise, a correction
leading to the \textsc{qed} description as shown above). Secondly, the
quantum description (according to \textsc{lsed}) refers to a kind of average
behavior, as follows from the principles of the theory and is amplified
here, and thus its dynamical variables are normally partially averaged
quantities describing the behavior of subensembles that comply locally with
the statistical requirements. A further but fundamental reason for the
appearance of sharp values for some dynamical variables is that they
correspond to stable stationary motions, which makes them emerge through the
statistics as preferred motions, selected by Principle Three. This is why
there can be eigenvalues at all in the theory. In the next subsections we
expand on these considerations.

The present formulation is based on a Fourier development of the field on
the frequency $\omega $, which means that each term of the development
corresponds to an infinity of field components with all possible values for
the wave vector $\mathbf{k}$ and polarization, with $k=\omega /c.$ In other
words, in each case we are considering the combined effect of all such
stochastic components, which vary from realization to realization, as a
single, unique instance, a simplification that is equivalent to perform a
partial averaging over the corresponding field modes \cite{Dice}. This is
one of the reasons we stated above that the dynamic variables are frequently
partially averaged quantities. A second obvious reason of the said implicit
partial averaging is the neglect of the effects of the noisy
(``nonrelevant'') components of the field.

Since once the quantum regime is reached and Principle Three holds, the
detailed motions do not depend on the specific realization of the field, it
becomes impossible to trace back the trajectory followed by a given particle
that reached the corresponding stationary state. In this sense, the
description becomes independent of the initial conditions and refers only to
subensembles, i.e., the set of those particles that reached the final state,
whatever the trajectory they may have followed.

It seems convenient to further elaborate on the matter. We have just seen
that Principle Three, by selecting the reduced set of solutions that are
insensitive to the specific realization of the field and thus particularly
stable, is the source of quantization in the present theory. From a more
physical point of view it becomes intuitively clear that the demand of
detailed energy balance can be satisfied only by a reduced (and frequently
discrete) set of motions. As we have seen, this latter requirement is the
outcome of the very stringent conditions imposed by the simultaneous demand
of energy balance and independence of the response functions (and the
relevant frequencies) from the specific realization of the field (Principle
Three). It is the mutual reinforcement of these two requirements which leads
to the selection of a well defined class of stationary allowed solutions,
and thus to quantization. By its physical content, the present explanation
of quantization stands in sharp contrast to the usual one related with the
properties of the wave function, although it is \emph{formally} similar to
the one afforded by matrix mechanics, where quantization arises from solving
a set of simultaneous algebraic equations similar to those given by Eq.(\ref%
{55a}) (with due allowance for the normalization to avoid the numerator $e%
\widetilde{E}_{\alpha \beta },$ and neglecting the radiation reaction
force). However, this explanation lacks the transparency provided by
Principle Three, besides being purely formal. In the Appendix it is shown
that the allowed stationary solutions correspond to extremum (indeed,
minimum) values for the energy.

\subsection{Are there trajectories?}

This is a major point in the interpretations of quantum mechanics. Since the
notion of trajectory does not enter into the axioms of quantum mechanics,
the dominant point of view is that in this theory (as it is) there are no
trajectories. However, this correct conclusion is frequently amplified to
mean that in nature there are no trajectories, when one refers to the
systems dealt with in quantum mechanics. This conclusion is tightly bound to
the origins of quantum mechanics, for instance, to the foundational work of
Heisenberg \cite{Jammer}, and has found its way to almost every textbook on
the subject. The argument is founded on the Heisenberg inequalities related
to noncommuting variables, such as $x$ and $p.$ Now in the present theory,
the trajectories exist, as follows from the starting premises out of which
the quantum description emerges. However, as discussed above, the theory
does not strictly describe the motion of individual particles, but of
subensembles of particles that satisfy (approximately) the statistical
demands on which it is constructed. From such a description, the individual
trajectory becomes unrecoverable. Therefore, the trajectories exist in
nature (as accounted in the initial description), but they do not belong to
the set of ingredients that comprise the final (partially averaged,
approximate and time-asymptotic) description. Hence quantum mechanics cannot
legitimately be used as a weapon against realism, as is frequently done.

The absence of trajectories in the quantum description is a serious obstacle
for the description of fast events, as are the transitions between states,
the quantum jumps. In the usual description there is some magic in these,
since the jumping electron must `know' in advance the energy of the orbit to
which it will be landing, to decide the frequency of the photon to be
radiated or absorbed. According to the present theory, transitions occur due
to resonant interactions with the background field. Given the state of the
atomic electron, there is a defined set of resonant frequencies to which it
may respond. Which will be the one selected in each instance is a matter of
chance, but there is no more guessing by the part of the electron. Of
course, `chance' should be understood here to mean that the end result
depends, among other things, on the specific realization of the stochastic
vacuum field, upon which we have no control.

There are attempts to introduce hidden variables into quantum theory to
recover the hidden trajectories, the best known one being Bohm's causal
theory \cite{Bohm}. According to the present view such attempts are doomed
to failure, since the individual behavior of a particle becomes
irretrievable once its stochastic motion has been smoothed out, either by
averaging or by approximations. The only sensible way to follow the real
trajectories is to go back to the original equation of motion (\ref{11}),
but even then we have the intrinsic problem of any stochastic description,
namely the specific realization of the field is unknown and with it the
specific trajectory. The best we can do in any real situation is to resort
to a statistical treatment of the problem. In plain words, this means that
to the extent that \textsc{lsed} is a sensible theory, the mere addition of
hidden variables to the usual quantum mechanical description to recover
determinism or realism is a very poor course. Even if one attempts to
complete the theory by adding the background field (as is usually done in 
\textsc{qed}), the trajectory of a specific particle remains undetermined;
this simply means that an indeterministic description of the quantum system
is unavoidable. It is interesting to compare this with the old eagerness,
expressed so many times by Einstein as his most tenacious devotee, for a
final description free of statistical elements. Unfortunately (at least for
some) that seems to be untenable. It is convenient to stress once more that
this does not mean a noncausal behavior of the particle: the full theory is
both causal and realist, since it is a branch of electrodynamics.

This latter remark is conveniently supported by the work of Cole and Zou
already cited \cite{ColZou03}, where all calculations are performed by
following the particles along their trajectories and computing the relevant
probability distributions. As already noted, with such procedure the authors
recover results that are close to the quantum mechanical ones for the ground
state of the hydrogen atom. We have thereat an explicit numerical example of
the possibility of interpreting the quantum results in terms of trajectories 
\cite{note1b}.

\subsection{Single particle versus ensemble interpretation}

Here we are at the core of the problems of interpretation of quantum
mechanics, since the answer to the present dilemma in one or the other sense
defines the person as an orthodox or unorthodox (and thus heretic). The
present theory gives an answer to this quandary and just on the iconoclastic
side. A theory that satisfies Principle Three (and detailed balance as is
here defined) cannot lead to the detailed description of the motions of a
single particle, but gives in a natural way a statistical rendering as
discussed formerly. A single particle will almost never satisfy the
principles of the present theory, although a subensemble of a big enough
collection of similarly prepared systems can satisfy them statistically. It
cannot be excluded that under certain conditions a single particle follows
closely enough the principles of the theory; under such circumstance of
course the theory describes approximately a single particle. But one swallow
does not make a summer, so we must adhere to the ensemble interpretation
when trying to extract the general implications of the present theory. This
point of view is certainly reinforced by the remaining considerations in
this section and the whole of the paper.

The above conclusion allows us to remove the need for the observer and the
collapse of the state vector, thus avoiding the paradoxes that they entail 
\cite{note2}. The observer becomes unnecessary because different results in
a series of a given measurement performed on the \emph{same} ensemble are
the direct result of measuring on different members of the ensemble. The
reduction (or collapse) of the state vector becomes dispensable because the
realization of a measurement on one of the partners of an entangled system
means changing the ensemble to adjust it to the new knowledge afforded by
the result of the measurement, which is just equivalent to a reduction of
the state vector \cite{Home}.

\subsection{Quantum Fluctuations and Uncertainty}

Quantum fluctuations are usually considered irreducible, on the basis of
relationships of the kind $\Delta x\Delta p\geq \hbar /2.$ These
inequalities are often interpreted in terms of unavoidable perturbations
attributed to measurements or, in a hazier language, to observations. In the
present theory these fluctuations reappear, but now as a result of the
interaction of matter with the background stochastic field ---which in
quantum mechanics remains hidden until we appeal to \textsc{qed}---, and
they attain their full force only when the system reaches the quantum
regime. Thus, from the perspective of \textsc{lsed} the quantum fluctuations
are not intrinsic to matter, but induced upon it by its interaction with the
vacuum field. Since the vacuum fluctuations are measured by $\hbar $, as
shown by Eq.(\ref{105}), also the induced equilibrium fluctuations, as
measured by relations such as $\Delta x\Delta p\geq \hbar /2,$ are
determined by $\hbar . $ This not only fixes the general scale of the
(minimum) fluctuations, but gives a causal meaning to them. Thus \textsc{lsed%
} implies that one should consider the quantum properties of matter not as
intrinsic (hence irreducible), but as acquired properties. It is important
to observe that, according to the present theory, the Heisenberg
inequalities hold only once the quantum regime is attained. So for extremely
short time intervals after the particle gets connected to the vacuum, they
could be violated, and although this possibility is extremely difficult to
verify for the moment, it remains open in principle. For example, for an
oscillator of frequency $\omega _{0}$ the relaxation time is of order $(\tau
\omega _{0}^{2})^{-1},$ which, for optical frequencies, is about 10$^{-10}$
s.

\subsection{Quantum non-causality and indeterminism}

As noted above, for \textsc{lsed} an entirely deterministic description \cite%
{note3} of the behavior of a quantum system seems to be an impossible task.
However, this is neither the result of the perturbations of the system by
our observations nor even less is it due to an intrinsic, ontological
indeterminism of the electron, as is usual to assume. Such behavior is
simply the result of the electron being in constant contact with a
stochastic ---thus unknown--- field. Had we strictly adhered to the detailed
original description (instead of developing an approximate statistical
formulation), and assumed the field to be known, everything would remain
causal and determined. But the kind of system we are considering and the
approximations and restrictions made along the derivation of the main
equations leading to \textsc{lsed} contravene these requirements, and so the
ensuing theory violates in principle both causality and determinism.
Causality is lost, since the agent responsible for the quantum (fluctuating)
behavior of matter, the vacuum field, is neglected in the quantum mechanical
description. As already said, a partial restitution of causality is achieved
in the transition from quantum mechanics to \textsc{qed}, but it is
introduced too late to recover a fully causal theory.

There are other instances in theoretical physics where approximations
transform an otherwise causal theory into one that violates causality.
Perhaps one of the best known examples is the Abraham-Lorentz equation of
motion (also used here). This equation is derived from a perfectly causal
combination of Maxwell's theory and classical mechanics. The end result, the
Abraham-Lorentz equation, can however give rise to noncausal phenomena as
preacceleration, the anticipated response to a future force. Again in this
case, the root of such noncausal behavior is to be found in the
approximations leading from the original causal and full description to the
final simplified (and noncausal) one. Approximate physical theories are not
bound to satisfy the same rigorous requirements that fundamental theories
are supposed to fulfil; this is particularly true with regard to consistency
with first principles \cite{Dice}.

\subsection{Wavelike behavior of matter}

\textsc{Lsed} contains a physical field in interaction with matter, and thus
it should be able to explain the appearance of the wave behavior of matter
as something not intrinsic, but impressed by the field and revealed by the
particles. This idea has been a guiding element of \textsc{sed} \cite{Boy75a}%
, \cite{Tiw86}, \cite{FeSa94} for a long time and is worth closer attention,
because it helps to develop a heuristic picture of some of the most puzzling
properties of quantum systems \cite{KrCdlP}. Our point of departure here is
that the linear response to the field characteristic of \textsc{lsed,} means
that where several fields superpose, the quantum response functions will
add. Thus, the degree of coherence of the underlying superposed fields will
be reflected under appropriate conditions in the coherence of the `guided'
matter, so to speak.

For the purpose of illustration, let us recall the typical example of the
double slit setup, with the detector far away from the two slits. Quantum
mechanics tells us that the passage of an electron through one slit is
affected by the existence of the neighboring slit, but it gives no physical
explanation to this fact. Of course once more we know the formal answer,
that such behavior is a consequence of the superposition of probability
amplitudes. We have at hand also the popular `explanation' that the
phenomenon is due to the self-interference of the electron. But strictly
speaking, this explains nothing, it merely describes what we observe.
Neither the formal answer nor the popular one solves the puzzle, which in
his famous lectures \cite{FLS} Feynman considered as the real mystery of
quantum mechanics. How is it that the mere existence of a second slit
affects the passage of an electron through the other nearby slit?

A qualitative explanation to this question can be offered from the
perspective of \textsc{lsed}---to provide a quantitative answer remains an
open task, although some work on it is in course. Any nearby body modifies
the background field, so that in the neighborhood of a periodic structure
the components of the field that fall onto it are enhanced in some preferred
directions due to diffraction, and curtailed in others. Under the knowledge
that the electron responds more strongly to the relevant waves of the
zero-point field, the main effect of diffraction on the particle will be to
reinforce the angular deviations specific to such waves, thus giving shape
to an interference pattern superimposed to the noisy background. Hence it is
the background field that carries the required information and operates
accordingly \textit{on} the particle. The picture that emerges reminds us of
the image suggested by J. Clauser some time ago: ``If a bunch of surfers
pass through a breakwater with two entrances, you'll see the two-slit
pattern later on the beach in surfer flesh!''\ (quoted in \cite{DW}, p.
116). And indeed, we have been observing for over 70 years many-slit
patterns in electron flesh.

\subsection{Final remark}

We have found that \textsc{lsed} explains in a most natural way some basic
properties that distinguish quantum systems from the corresponding classical
ones, including the wave-like properties of matter, atomic stability,
quantization, indeterminism, and so on, in addition to leading to the
correct quantitative description. Despite these most favourable traits that
substantiate its postulates, the theory here disclosed contains several
insufficiencies, the most important among them being the lack of an
assessment of the probability with which the trajectories can meet Principle
Three within reasonable limits. This is equivalent to an evaluation of the
probabilities with which the original free field amplitudes would evolve
towards the amplitudes $a_{\alpha \beta }$ that fulfil Eq.(\ref{52}).with
reasonable accuracy and within acceptable time intervals. This is a primary
problem that requires close scrutiny to confirm the soundness of the present
theory.

\section{Appendix}

In this Appendix we show that the mean value of the energies associated with
the solutions that comply with the principles of \textsc{lsed} correspond to
an extremum, in fact a minimum. We start by analyzing the mean kinetic
energy as follows from the solutions given by Eq.(\ref{23a}), which may be
written in the following form, introducing all the required indexes, but
neglecting the Larmor term,%
\begin{equation}
\left\langle T\right\rangle =\frac{1}{2m}\left\langle p^{2}\right\rangle =%
\frac{m}{2}\sum_{\beta \beta ^{\prime }}\omega _{_{\alpha }\beta }\omega
_{\alpha \beta ^{\prime }}\widetilde{x}_{\alpha \beta }^{\ast }\widetilde{x}%
_{\alpha \beta ^{\prime }}\left\langle a_{\alpha \beta }^{\ast }a_{\alpha
\beta ^{\prime }}\right\rangle e^{-i\left( \omega _{_{\alpha }\beta }-\omega
_{\alpha \beta ^{\prime }}\right) t}.  \tag{A1}
\end{equation}%
We consider a small variation due the independent variation of the
amplitudes $a_{\alpha \beta }^{\ast },a_{\alpha \beta ^{\prime }}:$%
\begin{equation}
\delta \left\langle T\right\rangle =\frac{m}{2}\sum_{\beta \beta ^{\prime
}}\omega _{_{\alpha }\beta }\omega _{\alpha \beta ^{\prime }}\widetilde{x}%
_{\alpha \beta }^{\ast }\widetilde{x}_{\alpha \beta ^{\prime }}\left\langle
a_{\alpha \beta }^{\ast }\delta a_{\alpha \beta ^{\prime }}+a_{\alpha \beta
^{\prime }}\delta a_{\alpha \beta }^{\ast }\right\rangle e^{-i\left( \omega
_{_{\alpha }\beta }-\omega _{\alpha \beta ^{\prime }}\right) t}.  \tag{A2}
\end{equation}%
We are considering that Principle Three holds, whence neither $\widetilde{x}%
_{\alpha \beta }$ nor $\omega _{_{\alpha }\beta }$ depends any more on the
amplitudes $a_{\alpha \beta }.$ Now from Eq.(\ref{50}) it follows that%
\begin{equation}
\delta a_{\lambda \mu }=ia_{\lambda \mu }\delta \phi _{\lambda \mu
}=ia_{\lambda \mu }\left( \delta \varphi _{\lambda }-\delta \varphi _{\mu
}\right) .  \tag{A3}
\end{equation}%
Since $\varphi _{\alpha }$ is common to both $a_{\alpha \beta }$ and $%
a_{\alpha \beta ^{\prime }}$ the independence of their variation means that
only $\varphi _{\beta }$ and $\varphi _{\beta ^{\prime }}$ change, so that $%
\delta a_{\alpha \beta ^{\prime }}=-ia_{\alpha \beta ^{\prime }}\delta
\varphi _{\beta ^{\prime }},$ $\delta a_{\alpha \beta }^{\ast }=ia_{\alpha
\beta }^{\ast }\delta \varphi _{\beta }$. Therefore,%
\begin{equation*}
\delta \left\langle T\right\rangle =i\frac{m}{2}\sum_{\beta \beta ^{\prime
}}\omega _{_{\alpha }\beta }\omega _{\alpha \beta ^{\prime }}\widetilde{x}%
_{\alpha \beta }^{\ast }\widetilde{x}_{\alpha \beta ^{\prime }}\left\langle
a_{\alpha \beta }^{\ast }a_{\alpha \beta ^{\prime }}\right\rangle \left(
\delta \varphi _{\beta }-\delta \varphi _{\beta ^{\prime }}\right)
e^{-i\left( \omega _{_{\alpha }\beta }-\omega _{\alpha \beta ^{\prime
}}\right) t} 
\end{equation*}%
\begin{equation*}
=i\frac{m}{2}\sum_{\beta \beta ^{\prime }}\omega _{_{\alpha }\beta }\omega
_{\alpha \beta ^{\prime }}\widetilde{x}_{\alpha \beta }^{\ast }\widetilde{x}%
_{\alpha \beta ^{\prime }}\left\langle a_{\alpha \beta }^{\ast }a_{\alpha
\beta ^{\prime }}\right\rangle \left( \delta \varphi _{\beta }-\delta
\varphi _{\beta ^{\prime }}\right) e^{-i\left( \omega _{_{\alpha }\beta
}-\omega _{\alpha \beta ^{\prime }}\right) t} 
\end{equation*}%
\begin{equation}
=i\frac{m}{2}\sum_{\beta \beta ^{\prime }}\omega _{_{\alpha }\beta }\omega
_{\alpha \beta ^{\prime }}\widetilde{x}_{\alpha \beta }^{\ast }\widetilde{x}%
_{\alpha \beta ^{\prime }}\delta _{\beta \beta ^{\prime }}\left( \delta
\varphi _{\beta }-\delta \varphi _{\beta ^{\prime }}\right) e^{-i\left(
\omega _{_{\alpha }\beta }-\omega _{\alpha \beta ^{\prime }}\right) t}=0. 
\tag{A4}
\end{equation}%
Due to the properties of the amplitudes $a_{\lambda \mu }$ under Principle
Three, a similar result holds for $\left\langle V(x)\right\rangle ,$ as
follows from a power series expansion. Thus we conclude that $\left\langle
\delta E\right\rangle =0$ to first order under the principles of \textsc{lsed%
} for arbitrary independent variations of the phases of the stochastic
amplitudes. This verifies that the mean energy of the stationary solutions
corresponds to an extremum. Moreover, it is clear that these extrema are
indeed minima, since otherwise the states would become unstable. To this
last observation we can arrive in a simpler way by recalling that the
answers afforded by the theory are just the quantum mechanical ones, which
very often correspond to the solutions of an eigenvalue problem. It is well
known that the eigenvalues of hermitian operators are local minima
determined by a variational principle. This is of course the case of the
energy eigenvalues, which shows that indeed the extremum values of the
energies $E_{\alpha }$ correspond to minima.

The observation that Eq.(\ref{104}) is identically satisfied at each
relevant frequency with a vacuum density $\rho \sim \omega ^{3},$ as given
in Eq.(\ref{105}) (for $T=0;$ for higher temperatures the same will hold for
the Planck spectrum), although obvious from the present point of view is
however highly nontrivial, since for classical separable systems, for
example, \ equilibrium occurs only with the Rayleigh-Jeans spectrum,
proportional to $\omega ^{2}$ \cite{vVH}.


\begin{thebibliography}{99}
\bibitem{Aul} G. Auletta, \emph{Foundations and Interpretation of Quantum
Mechanics }(World Scientific, Singapore, 2000).

\bibitem{Bohm} See e.g. D. Bohm and B. J. Hiley, \emph{The Undivided Universe%
} (Routledge, London, 1993).

\bibitem{Omnes} Omn\`{e}s \emph{The Interpretation of Quantum Mechanics}
(Princeton University Press, Princeton, New Jersey, 1994).

\bibitem{Jam} M. Jammer, \emph{The Philosophy of Quantum Mechanics. The
Interpretations of Quantum Mechanics in Historical Perspective }(John Wiley
and Sons, New York, 1974).

\bibitem{Mar63} T. W. Marshall, Proc. Roy. Soc. A\textbf{276}, 475 (1963);
Proc. Camb. Phil. Soc. \textbf{61}, 537 (1965).

\bibitem{MarSanVid94} T. W. Marshall, E. Santos and A. Vidiella-Barranco, in 
\emph{Proceedings of the Third International Workshop on Squeezed States and
Uncertainty Relations}, D. Han, Y. S. Kim, H. Rubin, Y. Shih and W. W.
Zachary (eds.), NASA Conference Publication Series no. 3270, NASA, 1994, p.
581.

\bibitem{MarSan02} T. W. Marshall and E. Santos, Recent Res. Devel. Optics 
\textbf{2}, 683 (2002).

\bibitem{Dice} L. de la Pe\~{n}a and A. M. Cetto, \emph{The Quantum Dice. An
introduction to Stochastic Electrodynamics} (Kluwer Acad. Publ., Dordrecht,
1996).

\bibitem{Boy70c} See, e.g., Th. H. Boyer, Annals of Physics \textbf{56}, 474
(1970).

\bibitem{Boy75a} Th. H. Boyer, Phys. Rev. D \textbf{11}, 790 (1975).

\bibitem{Boy80a} Th. H. Boyer, Phys. Rev A \textbf{21,} 66 (1980); see also
Th. H. Boyer, in \emph{Foundations of Radiation Theory and Quantum
Electrodynamics}, A. O. Barut (ed.) (Plenum Press, London, 1980).

\bibitem{dlPJau83} L. de la Pe\~{n}a and A. J\'{a}uregui, J. Math. Phys. 
\textbf{24}, 2751 (1983).

\bibitem{cavity} A. M. Cetto and L. de la Pe\~{n}a, Phys. Rev. A \textbf{37}%
, 1960 (1988); 1952 (1988).

\bibitem{Boy84} Th. H. Boyer, Phys. Rev. D \textbf{29}, 1089 (1984); D 
\textbf{30}, 1228 (1984). D. C. Cole, Phys. Rev. D \textbf{31}, 1972 (1985);
D \textbf{35}, 562 (1987).

\bibitem{Col85} D. C. Cole, Phys. Rev. D \textbf{31}, 1972 (1985).

\bibitem{Rue90} A. Rueda, Space Sc. Reviews \textbf{53}, 223 (1990).

\bibitem{San74} E. Santos, Nuovo Cim. B \textbf{19}, 57 (1974).

\bibitem{dlPCet79} L. de la Pe\~{n}a and A. M. Cetto, J. Math. Phys. \textbf{%
20}, 469 (1979).

\bibitem{FraMar88} H. M. Fran\c{c}a and T. W. Marshall, Phys. Rev. A \textbf{%
38}, 3258 (1988).

\bibitem{Boy76} Th. H. Boyer, Phys. Rev. D \textbf{13}, 2832 (1976); Phys.
Rev. A \textbf{18}, 1228 (1978).

\bibitem{ClaSot82} P. Claverie and F. Soto, J. Math. Phys. \textbf{23}, 753
(1982).

\bibitem{MarCla80} T. W. Marshall and P. Claverie, J. Math. Phys. \textbf{21}%
, 1819 (1980).

\bibitem{PesCla82} L. Pesquera and P. Claverie, J. Math. Phys. \textbf{23},
1315 (1982).

\bibitem{San85} E. Santos, in \emph{Stochastic Processes Applied to Physics}%
, L. Pesquera and M. A. Rodr\'{\i}guez (eds.) (World Scientific, Singapore,
1985); R. Blanco, L. Pesquera and E. Santos, Phys. Rev. D \textbf{27}, 1254
(1983); D \textbf{29}, 2240 (1984).

\bibitem{Col90} D. C. Cole, Found. Phys. \textbf{20}, 225 (1990).

\bibitem{ColZou03} D. C. Cole and Y. Zou, Phys. Letts. A \textbf{317}, 14
(2003); Phys. Rev. E \textbf{69 }016601 (2004); J. Sci. Computing \textbf{21}%
, 145 (2004) and references therein. See comments by T. H. Boyer, Found.
Phys. Lett. \textbf{16}, 613 (2003) and by P. W. Milonni, Found. Phys. Lett. 
\textbf{16}, 619 (2003).

\bibitem{elaf} L. de la Pe\~{n}a and A. M. Cetto, Linear Stochastic
Electrodynamics: Looking for the Physics Behind Quantum Theory, in \emph{New
Perspectives on Quantum Mechanics}, S. Hacyan, R. J\'{a}uregui and R. L\'{o}%
pez-Pe\~{n}a (eds.), AIP Conference Proceedings 464, New York, 1999. There
is an unabridged version of this work: \emph{Stochastic Electrodynamics:
Looking for the Physics Behind Quantum Theory}. Course given at the XXXI
Latin American School of Physics (ELAF), July 1998 (102 pages), that can be
requested from the authors.

\bibitem{dlPCet01} L. de la Pe\~{n}a and A. M. Cetto, Found. Phys. \textbf{31%
}, 1703 (2001).

\bibitem{note0} Distinct suggestions with similar purposes were given by
other authors also; see e.g. Th. H. Boyer, Found. Phys. \textbf{19}, 1371
(1989); D. C. Cole, Found. Phys. \textbf{20}, 225 (1990).

\bibitem{Ein16} A. Einstein, Mitteilungen Physikal. Ges. Z\"{u}rich, No. 18,
1916; reprinted in Phys. Zs. \textbf{18}, 121 (1917). English translation in
D. ter Haar, \emph{The Old Quantum Theory} (Pergamon, Oxford, 1967) and in
B. L. van der Waerden, \emph{Sources of Quantum Mechanics} (North Holland,
Amsterdam, 1967 and Dover, New York, 1968). There is also a translation into
French with commentaries in \emph{Albert Einstein, Oeuvres Choisies 1, Quanta%
} (Editions du Seuil-CNRS, Paris, 1989).

\bibitem{rho} That the energy of each field mode should be proportional to
its frequency is a relativistic demand, equivalent to asserting that the
spectral energy density of the vacuum field is proportional to $\omega ^{3}.$
This spectrum is the only one consistent with the demand, among others, of
being isotropic in all inertial systems. See, e. g., Ref. (\cite{Mar63}), T.
H. Boyer, Phys. Rev. \textbf{182}, 1374 (1969) or the discussion and ample
list of references in (\cite{Dice}).

\bibitem{Papo} A. Papoulis, \emph{Probability, Random Variables, and
Stochastic Processes} (McGraw-Hill, Tokyo, 1965).

\bibitem{BJH} M. Born, W. Heisenberg and P. Jordan, \emph{Zeitschr. f. Phys}%
. \textbf{35}, 557 (1926). Reprinted in \emph{Sources of Quantum Mechanics},
B. L. van der Waerden, ed. (Dover, New York, 1968).

\bibitem{Cohen} J. Dalibard, J. Dupont-Roc and C. Cohen-Tannoudji, J.
Physique \textbf{43}, 1617 (1982).

\bibitem{dlpc86} L. de la Pe\~{n}a and A. M. Cetto, Nuovo Cim. B \textbf{92}%
, 189 (1986).

\bibitem{ES83b} E. Santos, \emph{A definition of commutator of two
stationary processes. Application to stochastic electrodynamics}, preprint
1983, University of Santander (Spain), unpublished.

\bibitem{ES2} See e.g. E. Santos, J. Math. Phys. \textbf{15}, 1954 (1974);
also in \emph{Proceedings of the Einstein Centennial Symposium on
Fundamental Physics}, S. M. Moore, A. M. Rodr\'{\i}guez-Vargas, A. Rueda and
G. Violini (eds.) (Universidad de los Andes, Bogot\'{a}, 1981).

\bibitem{vVH} See e.g. J. H. van Vleck and D. L. Huber, Rev. Mod. Phys. 
\textbf{49}, 939 (1977).

\bibitem{note1} Due to the method of calculation used here there is no
difficulty in identifying the source and meaning of each of the terms
involved in Eqs. (\ref{115b}) and (\ref{115c}). This is not the case in the
usual \textsc{qed} treatments of the problem, because their interpretation
depends on the order of the creation and annihilation operators describing
the quantized field, and there exist a continuous number of combinations of
both possible orders. However, a detailed analysis of the problem leads just
to the same conclusion as in the text. A detailed account can be find in
Ref. \cite{Milonni}.

\bibitem{Milonni} See e.g. P. W. Milonni, \emph{The Quantum Vacuum. An
Introduction to Quantum Electrodynamics} (Academic Press, Inc., San Diego,
1994).

\bibitem{Lamb} W. E. Lamb and M. O. Scully in \emph{Polarisation, Mati\`{e}%
re et Rayonnement} (Presses Universitaires de France, Paris, 1969).

\bibitem{Jammer} See e. g. M. Jammer, \emph{The Conceptual Development of
Quantum Mechanics} (McGraw-Hill, New York, 1966).

\bibitem{note1b} In Bohm's causal interpretation there are also
trajectories. However, there is a fundamental difference between Bohm's
description and the trajectories of the present and Cole and Zou theories.
In the former theory the trajectories are constructed ad hoc from the
quantum results, whereas in both latter cases quantum mechanics follows from
the trajectories.

\bibitem{note2} We have here in mind the usual (and nonlocal) sense of the
term `collapse' in quantum theory, according to which a measurement
performed on one particle that belongs to an entangled state collapses the
state vector, so that both partners acquire well defined values for the
measured observable, independently of the distance between them. On occasion
the term is used to refer to real (and local) physical processes produced by
real interactions, as when photons from an entangled state pass through a
polarizer, so that those that pass become polarized and their state vector
is reduced by the physical interaction. We are excluding this second meaning
from our discussion.

\bibitem{Home} See e. g. D. Home, \emph{Conceptual Foundations of Quantum
Physics. An Overview from Modern Perspectives} (Plenum Press, New York,
1997).

\bibitem{note3} We are using the term \emph{deterministic} to refer to the
description, not to an ontological property, as is done and explained in
Ref. (\cite{Dice}) or in T. Brody, \emph{The Philosophy Behind Physics}, L.
de la Pe\~{n}a and P. Hodgson, eds. (Springer-Verlag, Berlin, 1993).

\bibitem{Tiw86} S. C. Tiwari, Proc. Einstein Found. Intern.\textbf{\ 3}, 63
(1986).

\bibitem{FeSa94} M. Ferrero and E. Santos, in \emph{Waves and Particles in
Light and Matter}, A. van der Merwe and A. Garuccio, eds. (Plenum, New York,
1994).

\bibitem{KrCdlP} For a complementary point of view see A. F. Kracklauer,
Phys. Essays \textbf{5}, 226 (1992); L. de la Pe\~{n}a and A. M. Cetto,
Found. Phys. \textbf{24}, 917 (1994), \textbf{25}, 573 (1995) or Ref. (\cite%
{Dice}).

\bibitem{FLS} R. P. Feynman, R. B. Leighton and M. Sands, \emph{The Feynman
Lectures in Physics,} Vol. \textsc{iii} (Addison-Wesley, Reading, MA, 1965).

\bibitem{DW} D. Wick, \textit{The Infamous Boundary. Seven Deades of Heresy
in Quantum Physics} (Copernicus, New York, 1995).

\newpage
\end{thebibliography}
\end{document}